\documentclass{ar2e}

\newcommand{\sqcm}{${\rm cm^{-2}}$}
\newcommand{\cubcm}{${\rm cm^{-3}}$}
\newcommand{\mum}{${\rm \mu m}$}
  
\newcommand{\waven}{${\rm cm^{-1}}$}

\newcommand{\av}{$A_{\rm V}$}

\usepackage{ulem}   
\usepackage{url}    
\usepackage{natbib} 
\usepackage{longtable} 
\usepackage{lscape} 
\usepackage{xcolor} 
\begin{document}

%


\input epsf.def   

\input psfig.sty

\jname{Annu. Rev. Astron. Astrophys.}
\jyear{2015}
\jvol{53}
\ARinfo{Boogert, Gerakines, \& Whittet}

\title{Observations of the Icy Universe}

\markboth{Boogert, Gerakines, \& Whittet}{Observations of the Icy
  Universe; ARAA 53 (2015; accepted), v. 05/05/2015}


\author{A. C. Adwin Boogert
\affiliation{Universities Space Research Association, Stratospheric
  Observatory for Infrared Astronomy, NASA Ames Research Center, MS
  232-11, Moffett Field, CA 94035, USA (acaboogert@alumni.caltech.edu)}
Perry A. Gerakines
\affiliation{Astrochemistry Laboratory, Solar
   System Exploration Division, NASA Goddard Space Flight Center,
   Greenbelt, MD 20771, USA}
Douglas C. B. Whittet
\affiliation{Department of Physics, Applied Physics and Astronomy and
  New York Center for Astrobiology, Rensselaer Polytechnic Institute,
  110 Eighth Street, Troy, NY 12180, USA}}


\begin{abstract}
Freeze-out of the gas phase elements onto cold grains in dense
interstellar and circumstellar media builds up ice mantles consisting
of molecules that are mostly formed in situ (H$_2$O, NH$_3$, CO$_2$,
CO, CH$_3$OH, and more).  This review summarizes the detected infrared
spectroscopic ice features and compares the abundances across
Galactic, extragalactic, and solar system environments. A tremendous
amount of information is contained in the ice band profiles.
Laboratory experiments play a critical role in the analysis of the
observations. Strong evidence is found for distinct ice formation
stages, separated by CO freeze out at high densities. The ice bands
have proven to be excellent probes of the thermal history of their
environment.  The evidence for the long-held idea that processing of
ices by energetic photons and cosmic rays produces complex molecules
is weak. Recent state of the art observations show promise for much
progress in this area with planned infrared facilities.

\end{abstract}

\maketitle

\section{PERSPECTIVE}

The condensation of gas phase elements on cold dust particles has long
been considered an important process in the Universe.  \citet{lin35}
studied the condensation of metals to form meteoritic material as the
starting point of planet formation. Considering the elemental
abundances, \citet{hul46} realized that the same process must lead to
the formation of H$_2$O, CH$_4$, and NH$_3$ ices on the grains.
\marginpar{\textcolor{blue}{\bf (Interstellar) ice:} solid state
  pure volatiles (H$_2$O, CO, etc.)  or mixtures thereof, with,
  perhaps, refractory inclusions, located on interstellar dust
  grains.}
Chemistry on cold grain surfaces has since emerged as a vital route to
formation of new molecules in dense clouds and circumstellar envelopes
and disks, including species that cannot readily be formed in the gas
phase (\citealt{tie82}; \citealt{her09} and references therein);
surface chemistry thus controls the composition not only of the icy
mantles but also of the molecular gas desorbed from the grains in the
vicinity of cloud edges, shocks, and the radiation fields of Young
Stellar Objects (YSOs).

Other chemical pathways to new interstellar ice species are
observationally not as well established.  Laboratory studies have
shown that ice heating creates new molecules by purely thermal
reactions \citep{sch93, the13} or through the increased mobility of
radicals created by ultraviolet (UV) photons and cosmic ray (CR)
particles. The latter has been a central topic in laboratory studies
as a means to form complex molecules from simple ices.  Heavy
processing of ices forms organic residues (sometimes referred to as
``yellow'' and ``brown stuff'') that resemble the insoluble organic
component of carbonaceous meteorites (\citealt{gre73, gre95}, and
references therein), and the products may include biologically
relevant species (e.g., \citealt{ber95, mun04}).

Ices are an important reservoir of the elements in dense ($>10^3$
\cubcm) environments; up to 60\% of the oxygen not included in
silicates is incorporated in known ice species.  The processes that
are key in their origin and evolution are thus directly relevant to
key questions on the organic and volatile reservoir of regions of star
and planet formation.  What is the origin of the volatiles and
organics of the solar nebula, possibly delivered to the early Earth as
the building blocks of life?
\marginpar{\textcolor{blue}{\bf Energetic processing of ices:}
  exposure to ionizing radiation, breaking bonds, and creating
  radicals and ions.  Subsequent chemical reactions may lead to more
  complex molecules.}  There is no clear consensus on whether these
materials survived the voyage from circumstellar envelopes into
protoplanetary disks (e.g., \citealt{vis09, bro14, cle14}).  Comets,
believed to contain the most pristine remnant ices from the formation
of the solar system (e.g., \citealt{mum11}), are key in answering this
question.

In the broader astrophysical context, ices play important roles as
well.  For example, they speed up the grain coagulation process
\citep{orm09}. Also, questions on the importance of ice processing are
directly related to questions about the origin, composition, and
physical properties of interstellar dust.  Energetically processed
ices might be a source of refractory interstellar dust, in particular
in star forming regions (\citealt{li14} and references therein) as
opposed to dust formed in the envelopes of evolved stars.

\marginpar{\textcolor{blue}{\bf Photodesorption:} non-thermal desorption of a particle
  from a grain surface after a photon hit to itself or an underlying
  particle.}  

The initial searches for ices targeted the 3.0 \mum\ H$_2$O band in
lines of sight dominated by diffuse dust \citep{dan65,
  kna69}. \citet{gil75} noted that the lack of ices in these diffuse
clouds could not be explained by elevated dust temperatures alone,
revealing the importance of the non-thermal process of photodesorption
\citep{wat72}.
The 3.0 \mum\ feature was finally detected toward the
Orion BN/KL region \citep{gil73}. Its position and width were in
reasonable agreement with a model of small, pure H$_2$O ice spheres,
but significant deviations were observed as well \citep{mer76}: short
and long-wavelength wings were attributed to scattering by large
grains and interactions with NH$_3$ mixed in the ice \citep{mer76,
  kna82, smi89}.  The latter supported the concept of ice mixtures
(sometimes called ``dirty ice'') first proposed by \citet{hul46}.
\marginpar{\textcolor{blue}{\bf Dirty ice:} ill-defined term sometimes used to
  indicate that interstellar ices are mixtures, contain processed
  residues, or are present on grains.}  Most ice features discovered
thereafter show stunningly complex profiles as a result of the ice
composition, structure (thermal history), and grain shape and
size. For several bands (those of H$_2$O, CO, and CO$_2$ in
particular) this is well understood, and their diagnostic value of the
physical conditions and the nature of astronomical sources is a main
topic of this review.

This is a review of observational studies of interstellar and
circumstellar ices.  The composition of the ices and evidence for
formation and thermal and energetic modification processes are
reviewed across Galactic and extragalactic environments.  Previous
reviews focused on massive YSOs (MYSOs hereafter; \citealt{gib04,
  boo04b, dar05}) and on low mass YSOs (LYSOs hereafter) and
background stars tracing nearby clouds \citep{obe11}.  In particular,
the study by \"Oberg et al. is expanded here to include the Galactic
Center (GC) region and extragalactic environments, as well as to a
full inventory of observed features and identifications.  Solar system
ices are not reviewed in detail, but the interstellar medium (ISM) and
circumstellar ice properties are compared with those of comets as
reviewed by \citet{mum11} to address the question of the origin of
solar system volatiles and organics.

\section{ICES ACROSS THE ELECTROMAGNETIC SPECTRUM}~\label{sec:em}

Ices in interstellar environments are traced almost exclusively by
their molecular vibrational transitions in the near- to
far-infrared. In general, the 1-3 \mum\ wavelength region contains the
combination and overtone modes, the 3-6 \mum\ region the stretch
vibrations, the 6-30 \mum\ region the bend and libration vibrations,
and the 25-300 \mum\ region the torsional and inter-molecular
(lattice) modes. Some ices also have strong electronic transitions in
the UV, e.g. in the case of H$_2$O near 0.14 \mum\ \citep{war08}.
These UV features have not yet been observed in the ISM, due at least
in part to very high UV extinction in candidate lines of sight.  They
may be detectable in absorption near cloud edges (at low \av\ values;
\citealt{goe83}) or through scattering. Finally, simple ices are
transparent outside their vibrational modes, but the continuum
opacities increase significantly for processed ices \citep{jen93,
  bru06} and ices with amorphous carbon inclusions \citep{pre93}.

\begin{figure}[t!]
\centerline{\psfig{figure=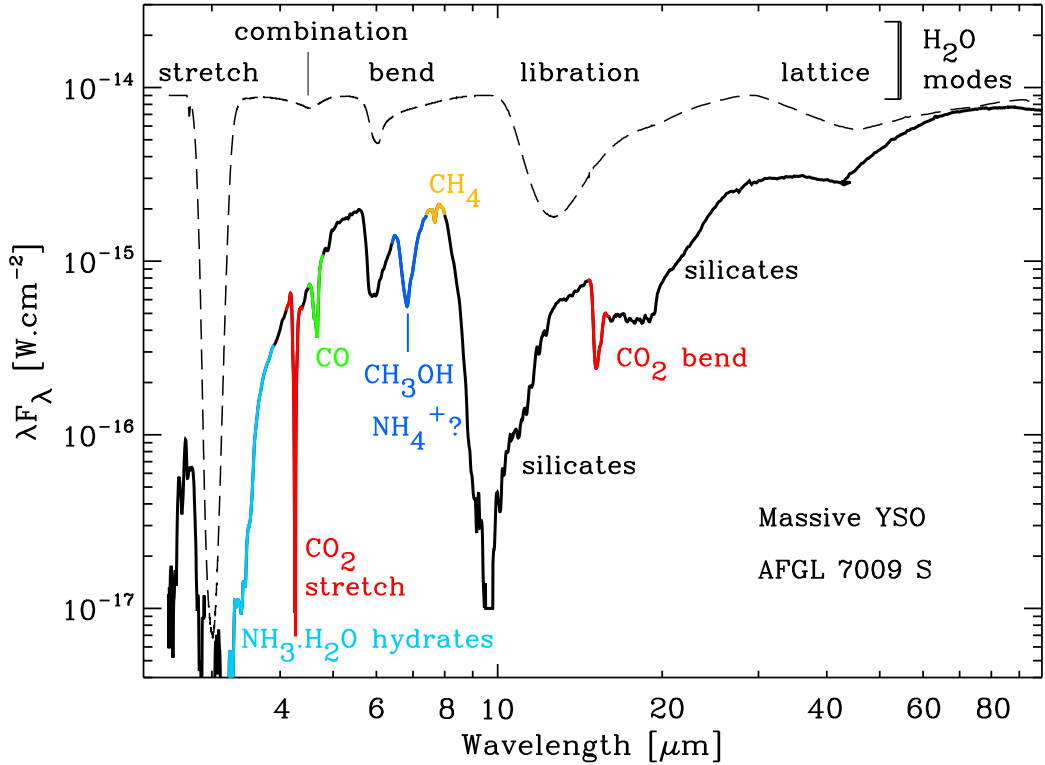,height=24pc,angle=90}}
\caption{Overview of the strongest ice and dust features in the MYSO
  AFGL 7009S \citep{dar98}. The calculated spectrum of pure H$_2$O ice
  spheres at 10 K is shown (dashed line) to indicate the multiple
  H$_2$O bands.}\label{f:globalice}
\end{figure}

\begin{figure}[h!]
\centerline{\psfig{figure=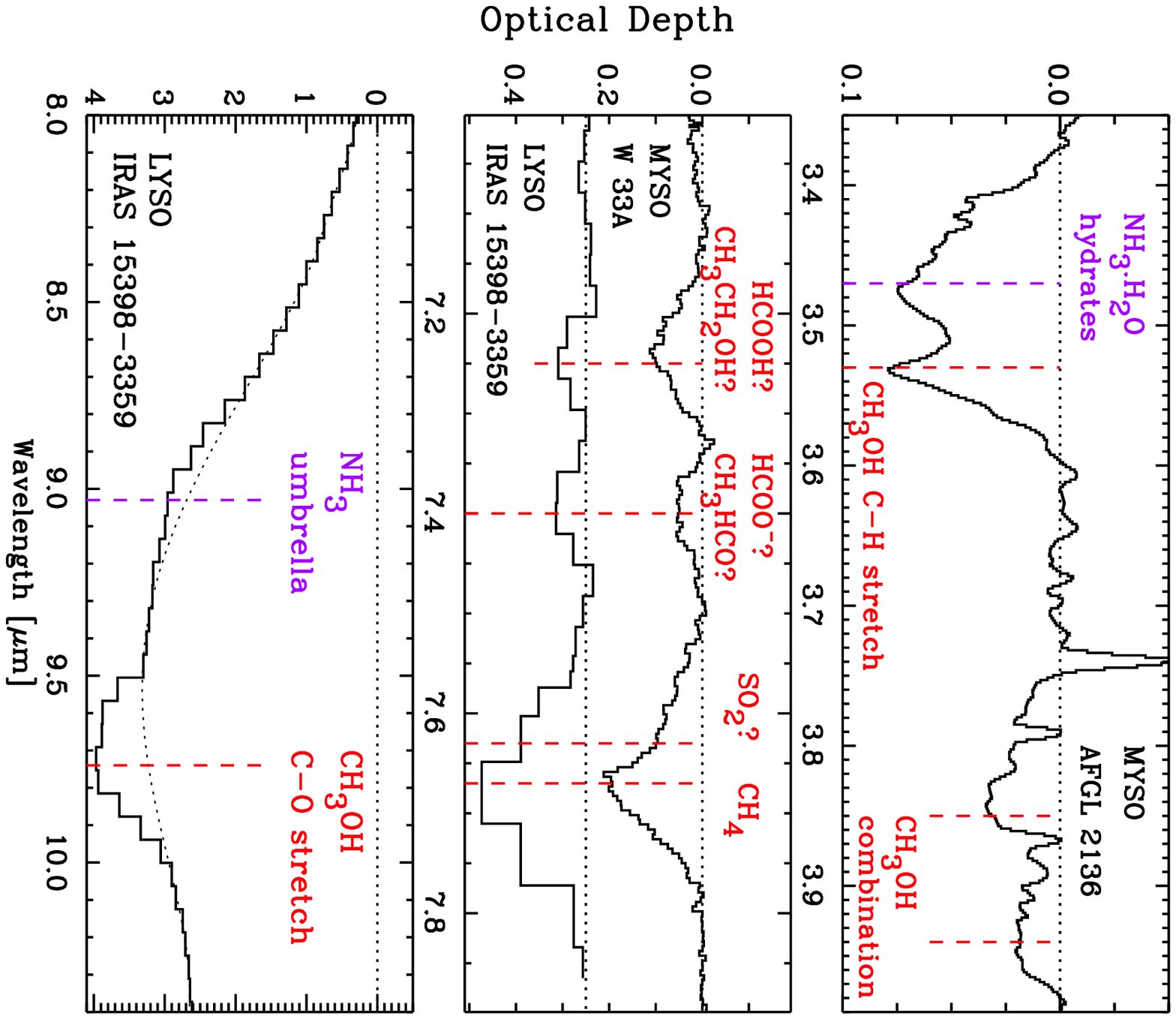,height=28pc,angle=90}}
\caption{Compilation of weak ice absorption features in the 3.3-4.0
  (top), 7.0-7.9 (middle) and 8.0-10.4 (bottom)
  \mum\ regions. Identifications are labeled, and tentative
  identifications are followed by question marks. Data for AFGL 2136
  were taken from \citet{dar03}, for W 33A from \citet{sch99}, and for
  IRAS 15398-3359 from \citet{boo08} and
  \citet{bot10}.}\label{f:weakice}
\end{figure}

The vast majority of interstellar and circumstellar ice studies target
the stretch and bend mode vibrations in the 3-16 \mum\ range, because
below 3 \mum\ the signals are weak as a result of dust continuum
extinction (in contrast to solar system studies which often observe
the combination modes at 1-2 \mum), and above $\sim$30 \mum\ the
availability and capability of instrumentation are limited.
Figures~\ref{f:globalice} and~\ref{f:weakice} show a selection of ice
features detected outside the solar system and Table~\ref{t:feat}
gives a comprehensive summary of the typical peak positions, widths,
and identifications.  Nearly all features are securely detected, but
not all are positively identified, as indicated with questions marks
and further discussed in \S\ref{sec:iden}.  The widths vary
dramatically, from 1.5 \waven\ (0.0034 \mum) for $^{13}$CO to 400
\waven\ (0.38 \mum) for the H$_2$O ice bands. Also, many band profiles
(e.g., those of CO and CO$_2$) contain substructures at the level of a
few \waven.  Thus, ice studies benefit significantly from the
availability of medium resolution infrared spectrometers
($R=\lambda/\Delta\lambda\geq$500).  For many features, the positions
and widths vary considerably across objects. These variations are
often well studied, and have led to empirical decompositions. The
components are listed, and are further discussed in \S\ref{sec:prof}.

Indirect constraints on the ice composition are provided by
observations of gas-phase rotational lines at (sub)-millimeter
wavelengths.  First, low gas phase abundances (depletions) of specific
molecules in clouds imply their presence in the ices
\citep{ber07}. Second, gas phase species observed in hot core regions
surrounding YSOs originate from sublimated ices, either directly or
after gas phase chemical reactions \citep{bla87, her09}. Third, some
gas near cloud edges originates from photodesorbed ices
\citep{obe09a}.  These results are not reviewed here, but are quoted
when relevant.

\subsection{Ice Mapping}\label{sec:map}

Most ice features are detected as pure absorption bands against
infrared continuum point sources. Studies of the spatial variations of
the ice properties are therefore relatively rare, yet very powerful.
They often rely on the presence of many point sources (usually
infrared-bright giants) behind clouds (e.g., Taurus;
\citealt{mur00}). The envelope of a Class 0 YSO was mapped using
background Class II LYSOs \citep{pon04}, and of a dense core using
Class I/II LYSOs \citep{pon06}. Ice mapping is however also possible
if the background emission is extended by scattering from dust in any
disk, envelope, or outflow cone \citep{har97, spo03, sch10}. In
addition, scattering by large grains enhances the short-wavelength
wing of the 3.0 \mum\ H$_2$O band (\S\ref{sec:30}; \citealt{pen90}),
which enables mapping of ices in reflection nebulae.  This has rarely
been done.  Similarly, the lattice modes ($\sim$ 25-300 \mum;
\S\ref{sec:lattice}) are suited for ice mapping. They are the only ice
bands that may appear in emission, because, in contrast to the
intra-molecular modes, their excitation energy is below the binding
energy of the ices.

\newpage
\vspace*{9cm}

\begin{landscape}
\begin{longtable}{lcclp{3cm}}
\caption{Ice Features Detected in Interstellar and Circumstellar
  Environments}\label{t:feat}\\
\toprule
$\lambda _{\rm center}$  & $\nu _{\rm center}$ & FWHM     & Identification  & Reference$^{\rm a}$ \\
$\mu$m                & cm$^{-1}$         & cm$^{-1}$ &                 &  \\
\colrule
2.27              &   4405          & 60           & CH$_3$OH combination                                & 1         \\
2.697             &   3708          & 4            & CO$_2$ combination                                  & 2         \\
2.778             &   3600          & 5            & CO$_2$ combination                                  & 2         \\
2.96              &   3378          & $\sim 40$    & NH$_3$ stretch?                                     & 3, 4    \\
3.0               &   3333          & $\sim$ 390   & H$_2$O stretch (multiple components)$^{\rm b}$        &      \\
---2.90           &   3448          & $\sim 120$   & --- scattering                                      & 5, 6     \\
---3.05           &   3278          & $\sim$ 315   & --- amorphous small grains                          & 7, 8     \\
---3.10           &   3225          & $\sim$ 220   & --- crystalline small grains                        & 8        \\
---$\sim$3.3      &   3030          & $\sim$ 270   & --- scattering and NH$_3$.H$_2$O hydrate            & 9, 10, 16     \\
3.25              &   3076          & 70           & PAH C-H stretch? NH$_4^+$ combination?              & 11, 12, 13 \\
3.325             &   3008          & 15           & CH$_4$ stretch                                      & 14        \\
3.47              &   2881          & 80           & NH$_3$.H$_2$O hydrate                               & 15, 4     \\
3.53              &   2832          & 30           & CH$_3$OH C-H stretch                                & 17         \\
3.85              &   2597          & $\sim 40$    & CH$_3$OH combination                                & 15         \\
3.94              &   2538          & $\sim 40$    & CH$_3$OH combination                                & 18, 15      \\
4.07$^{\rm c}$      &   2457          & $\sim 120$   & HDO O-D stretch amorphous                         & 19        \\
4.13$^{\rm c}$      &   2421          & $\sim 60$    & HDO O-D stretch crystalline                       & 19        \\
4.27              &   2341          & $\sim 20$    & CO$_2$ stretch                                      & 20        \\
4.39              &   2278          & 9            & $^{13}$CO$_2$ stretch (multiple components)         &         \\
---4.382          &   2282          & 3            & --- pure CO$_2$                                     & 21        \\
---4.390          &   2278          & 8            & --- polar, H$_2$O/CO$_2>$1                          & 21        \\
4.5               &   2222          & 270          & H$_2$O combination                                  & 22        \\
4.62              &   2164          & 29           & ``XCN'' (multiple components)                       & 23        \\
$\lambda _{\rm center}$  & $\nu _{\rm center}$ & FWHM     & Identification  & Reference$^{\rm a}$ \\
$\mu$m                & cm$^{-1}$         & cm$^{-1}$ &                 &  \\
\colrule
---4.598          &   2175          & 15           & ---OCN$^-$ C$\equiv$N stretch (apolar)?             & 24, 25   \\
---4.617          &   2166          & 26           & ---OCN$^-$ C$\equiv$N stretch (polar)               & 24, 25   \\
4.67              &   2141.3        & 3-9          & CO stretch (multiple components)                    & 26        \\
---4.665          &   2143.7        & 3.0          & --- apolar, CO$_2$/CO$>1$?                          & 44, 24    \\
---4.673          &   2139.9        & 3.5          & --- apolar, $>90\%$ CO                              & 23, 24    \\
---4.681          &   2136.5        & 10.6         & --- polar, CH$_3$OH/CO$>1$                          & 23, 24, 27\\
4.779             &   2092.4        & 1.50         & $^{13}$CO stretch (apolar)                          & 28        \\
4.90              &   2040          & 22           & OCS C-O stretch                                     & 18, 29     \\
5.83              &   1715          & 35           & H$_2$CO C-O stretch                                 & 30        \\
6.0               &   1666          & 160          & multiple components                                 & 31        \\
---6.0            &   1666          & $\sim 130$   & ---H$_2$O bend                                      & 31        \\
---5.84           &   1712          & 72           & ---HCOOH C-O stretch?                               & 30, 31, 32     \\
---6.18           &   1618          & 111          & ---NH$_3$ bend, H$_2$O/CO$_2$=2, PAH C-C stretch    & 30, 32     \\
6.0               & $\sim$ 1666     & $\sim 330$   & organic residue or an-ions?                      & 33, 32    \\
6.85              &   1459          & 85           & multiple components                                 & 31        \\
---6.755          &   1480          & 43           & ---CH$_3$OH C-H deformation, NH$_4^+$ bend?         & 30, 32    \\
---6.943          &   1440          & 61           & ---NH$_4^+$ bend?                                   & 30, 32    \\
7.24              &   1381          & 19           & HCOOH, CH$_3$CH$_2$OH, C-H deformation?             & 34        \\
7.41              &   1349          & 15           & HCOO$^-$, CH$_3$CHO, C-H deformation?               & 34        \\
7.63              &   1310          & 28           & SO$_2$ stretch?                                     & 35        \\
7.674             &   1303          & 11           & CH$_4$ deformation                                  & 36, 37    \\
8.865             &   1128          & 15           & CH$_3$OH C-H$_3$ rock                               & 38        \\
9.01              &   1109          & 50           & NH$_3$ umbrella                                     & 39        \\
9.74              &   1026          & 30           & CH$_3$OH C-O stretch                                & 38        \\
13.6              &    735          & $\sim 220$   & H$_2$O libration                                    & 40        \\
$\lambda _{\rm center}$  & $\nu _{\rm center}$ & FWHM     & Identification  & Reference$^{\rm a}$ \\
$\mu$m                & cm$^{-1}$         & cm$^{-1}$ &                 &  \\
\colrule
15.1-15.3         &    653-662      & 16-28        & CO$_2$ bend (multiple components)                   & 20        \\
---15.31          &    653          & 22           & --- polar, CO$_2$/H$_2$O$<1$                        & 24        \\
---15.10          &    662.2        & 11           & --- apolar, CO$_2$/CO$\sim$1                        & 24        \\
---15.15          &    660.0        & 2.5          & --- dilute, CO$_2$/CO$<<1$                          & 24        \\
---15.11+15.26    &  655.3+661      & 19           & --- pure CO$_2$                                     & 24        \\
---15.4           &    648          & 10           & --- shoulder, CH$_3$OH$/$CO$_2>>1$                  & 24        \\
44                &    227          & $\sim 100$   & H$_2$O lattice (amorphous)                          & 41, 42, 43\\
44                &    227          & $\sim 50$    & H$_2$O lattice (crystalline)                        & 41, 42, 43\\
62                &    161          & $\sim 50$    & H$_2$O lattice (crystalline)                        & 41, 42, 43\\
\botrule

\multicolumn{5}{p{17.5cm}}{$^{\rm a}$ reference to a publication
  contributing strongly to its identification or profile
  characterization:
  1:  \citealt{tab03},           
  2:  \citealt{kea01a},          
  3: \citealt{chi00},
  4: \citealt{dar01},
  5:  \citealt{kna82},          
  6: \citealt{pen90},
  7:  \citealt{gil73},    
  8: \citealt{smi89},
  9:  \citealt{mer76},         
  10: \citealt{kna87},
  11:  \citealt{sel94},        
  12: \citealt{har14},
  13: \citealt{sch03},
  14: \citealt{boo04a},
  15:  \citealt{all92},     
  16: \citealt{dar02},
  17:  \citealt{gri91},            
  18:  \citealt{geb85},         
  19: \citealt{aik12},
  20: \citealt{ger99},
  21: \citealt{boo00a},
  22: \citealt{boo00b},         
  23: \citealt{lac84},            
  24: \citealt{pon03},
  25: \citealt{bro05},
  26: \citealt{soi79},          
  27: \citealt{cup11},
  28: \citealt{boo02a},         
  29: \citealt{pal97},
  30: \citealt{kea01b},          
  31: \citealt{sch96},         
  32: \citealt{boo08},
  33: \citealt{gib02},
  34: \citealt{sch99},         
  35: \citealt{boo97},         
  36: \citealt{lac91},            
  37: \citealt{boo96},
  38: \citealt{ski92},         
  39: \citealt{lac98},            
  40: \citealt{cox89},                    
  41: \citealt{omo90},
  42: \citealt{mal99},
  43: \citealt{dar98},
  44: \citealt{boo02b}}\\

\multicolumn{5}{p{16cm}}{$^{\rm b}$ Contributions from the O-H stretch
  mode in CH$_3$OH and perhaps HCOOH and other species are expected,
  but are not detected as separate features.}\\

\multicolumn{5}{p{16cm}}{$^{\rm c}$ Tentative detection.}\\

\end{longtable}
\end{landscape}

\section{LABORATORY EXPERIMENTS}\label{sec:lab}

Laboratory experiments have been crucial in the identification of the
observed interstellar ice features (Table~\ref{t:feat}), in
understanding their profiles (\S\ref{sec:prof}), and in determining
the chemical origin and fate of their carriers (\S\ref{sec:model}). An
overview of laboratory work is beyond the scope of this review, but
key aspects will be emphasized here and mentioned throughout the
remaining sections.

The peak positions, widths, and shapes of the observed ice bands are
influenced by the composition and structure of the ices
(crystallinity, density, and homogeneity) as a result of the
attractive and repulsive dipole interactions between neighboring
molecules. They are also influenced by the properties of the dust
grains (size, shape, and the relative mantle and core volumes) and
these effects can be calculated from the optical properties of the
bulk materials. All of these influences must be taken into account in
the proper assignment of carriers to the observed ice features and
special care must be used in deriving physical and chemical
information from the band profiles such as overall ice composition or
temperature.

The absorption band profiles of ices and ice mixtures are routinely
measured by laboratory transmission and reflection spectroscopy at
high vacuum ($< 10^{-7}$ mbar) under temperature-controlled conditions
(e.g., \citealt{hud93}).  These laboratory samples may also be exposed
to UV photons or to particles, simulating molecule formation and
destruction by energetic processes in interstellar environments (e.g.,
\citealt{ger01b}).  Optical constants are generally measured with a
Kramers-Kronig analysis of a single spectrum using the ice's known
thickness and visible refractive index.  For CO and CO$_2$,
conflicting optical constants were reported \citep{ehr97}, and it was
shown that the most accurate values are derived using polarized light
at a range of sample thicknesses and incidence angles
\citep{bar98}. Databases of laboratory transmission spectra and
optical constants of interstellar ice analogs are listed in "Related
Resources".

Infrared band strengths (\S\ref{sec:abun}) can be determined directly
from the ice's complex refractive index, or from laboratory spectra of
an ice of a known thickness that must be determined independently.
The band strengths of a particular molecule vary with ice temperature
and overall ice composition, although measurements under astrophysical
conditions are relatively scarce.  The band strengths of CO and CO$_2$
were shown to vary by less than $\sim$20\% when measured at low
temperatures, mixed together, or in an H$_2$O ice matrix
\citep{ger95}, but the strength of the 3.0 \mum\ H$_2$O stretch mode
is reduced by an order of magnitude when H$_2$O is highly diluted
\citep{thi57, ehr96}. Integrated strengths of common ice bands are
listed in, e.g., \citet{hud93} and \citet{ger95}.

Improvements in laboratory techniques have made it possible to measure
fundamental parameters such as energy barriers, reaction rates, and
photodesorption yields. We refer to \citet{lin11} for a recent review
on this topic. Very briefly, this ``bottom up'' approach has been
successful in creating H$_2$CO and CH$_3$OH \citep{wat02}, CO$_2$
\citep{iop11} and other species on cold surfaces with quantitative
results. These have been used in grain surface-chemistry models and
have successfully reproduced observed ice abundances under realistic
astrophysical conditions (e.g., \citealt{cup11}).

\section{ICE BAND PROFILES}~\label{sec:prof}

The profiles of the 3.0, 4.67, and 15.2 \mum\ ice features have proven
to be sensitive probes of the physical conditions and the nature of a
wide range of astronomical objects (\S\ref{sec:env}): dense clouds,
Galactic and extragalactic YSOs, evolved stars, the Galactic Center
region, and external galaxies.  The lattice modes above 25 \mum\ show
great promise as probes of the ice composition and processing history.
The 6 and 6.8 \mum\ features, though not fully identified, were shown
to have diagnostic value as well.  All other bands are discussed in
\S\ref{sec:abun}.

\subsection{The 3.0 \mum\ H$_2$O Band: Crystallization, Large Grains, and Hydrates}\label{sec:30}

\begin{figure}[t!]
\centerline{\psfig{figure=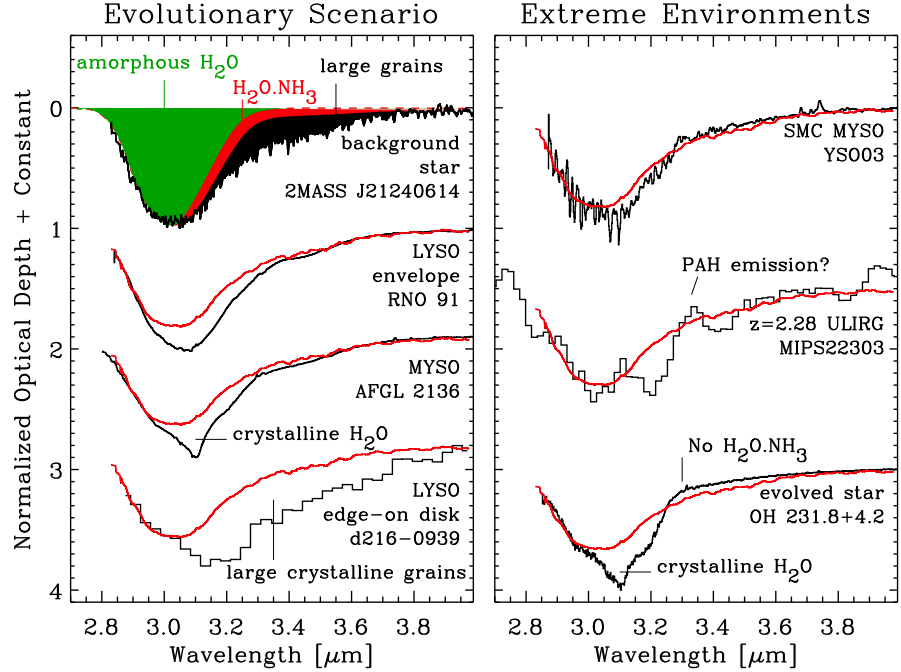,height=21pc,angle=90}}
\caption{Observed 3.0 \mum\ band profiles, highlighting the
  universally present H$_2$O ice components and their variations. The
  left panel shows sightlines with, from top to bottom, increasing
  degrees of crystallinity as a result of heating above temperatures
  of $\sim 70$ K: the background star 2MASSJ 21240614+4958310
  \citep{boo11}, the LYSO RNO 91 \citep{boo08}, the Galactic MYSO
  GL2136 \citep{dar02}, and the edge-on disk source d216-0939
  \citep{ter12a}. The right panel shows the SMC MYSO 03 \citep{oli13},
  the ULIRG galaxy MIPS 22303 at z=2.28 (\citealt{saj09}), and the
  evolved star OH 231.8+4.2 \citep{dar02}. For comparison, the
  smoothed spectrum of the background star is plotted over each
  spectrum (red).  The background star itself is compared with pure,
  amorphous H$_2$O ice particles (CDE shapes) at 10 K (green area). If
  instead a H$_2$O:NH$_3$=10:1 spectrum is used, part of the
  long-wavelength wing is explained by NH$_3$ hydrates (NH$_3$.H$_2$O;
  red area). The black area is the residual due to large ($\sim 0.35$
  \mum) grains. }\label{f:3um}
\end{figure}

The interstellar 3.0 \mum\ band (Fig.~\ref{f:3um}) is primarily due to
the O-H stretching mode of bulk H$_2$O ice, because of the high
abundance of this species and because of the very high intrinsic band
strength induced by the hydrogen bonding network \citep{thi57}). 
\marginpar{\textcolor{blue}{\bf Hydrogen bonding network:} dipole attraction between
  the electropositive H from one molecule with the electronegative
  atom (e.g., O or N) from another.}  The bands of H$_2$O ice mixed
with apolar species peak below 3.0 \mum, and, lacking the hydrogen
bonding network, their intrinsic strengths are an order of magnitude
smaller \citep{ehr96}. These features of isolated H$_2$O have not been
detected in interstellar environments.

The profile of the 3.0 \mum\ band is not fully explained by absorption
by small pure H$_2$O ice spheres \citep{mer76, kna87}.  Excess
absorption is always present between 3.3 and 3.7
\mum\ (Fig.~\ref{f:3um}), and sometimes near 2.9 \mum. The strengths
of the long wavelength wing and the peak depth of the 3.0 \mum\ band
correlate well. Some variations were reported
\citep{pon04,thi06,nob13}, suggesting a dependence on H$_2$O
abundance, but this needs further work.  In any case, Mie scattering
calculations show that large grains ($\sim 0.5$ \mum; \citealt{smi89};
\S\ref{sec:size}) might explain much but not all of the long
wavelength wing.  
\marginpar{\textcolor{blue}{\bf Ammonia hydrates:} a hydrogen bonding network between
  H$_2$O and NH$_3$, often noted as H$_2$O.NH$_3$.}
The remainder of the absorption is attributed to the
O-H stretching mode in ammonia hydrates \citep{kna82, dar01}.  The
distinct 3.47 \mum\ feature is likely also a result of these hydrates
(\S\ref{sec:iden}; \citealt{dar01}).

The wing at 2.9 \mum\ is not as commonly observed.  Its strength
increases with the observational beam size, capturing larger
contributions of light scattered off large ice-coated grains in some
objects (e.g., the Orion BN/KL nebula in \citealt{pen90}).  The
absence of this feature in a background star and its presence towards
YSOs and reflection nebulae further confirms this interpretation
\citep{smi89}.
\marginpar{\textcolor{blue}{\bf Thermal processing of ices:} increase in ice
  temperature, leading to sublimation, a rearrangement of molecular
  bonds (crystallization and segregation), or chemical reactions.}
The stretching mode of NH$_3$ contributes to this wing, but it is
rarely detected as a distinct feature \citep{chi00, dar01}.

The 3.0 \mum\ band profile varies considerably among targets within
and across different environment classes (Fig.~\ref{f:3um}). The peak
position and width toward the majority of observed YSOs and all
background stars are consistent with amorphous H$_2$O ice
\citep{smi89}. Narrower profiles that are shifted to longer
wavelengths toward some MYSOs and a few LYSOs \citep{boo08} reflect
higher degrees of crystallinity attributed to thermal processing
(\S\ref{sec:modeltherm}). An extreme case is that of the edge-on low
mass disk source d216-0939 \citep{ter12a}, which shows a peak at
$\sim$3.20 \mum\ caused by large ($\sim 0.8$ \mum) grains consisting
of nearly pure crystalline H$_2$O ice.

\subsection{The 4.67 \mum\ CO Band: the Most Volatile Ices}\label{sec:467}

\begin{figure}[t!]
\centerline{\psfig{figure=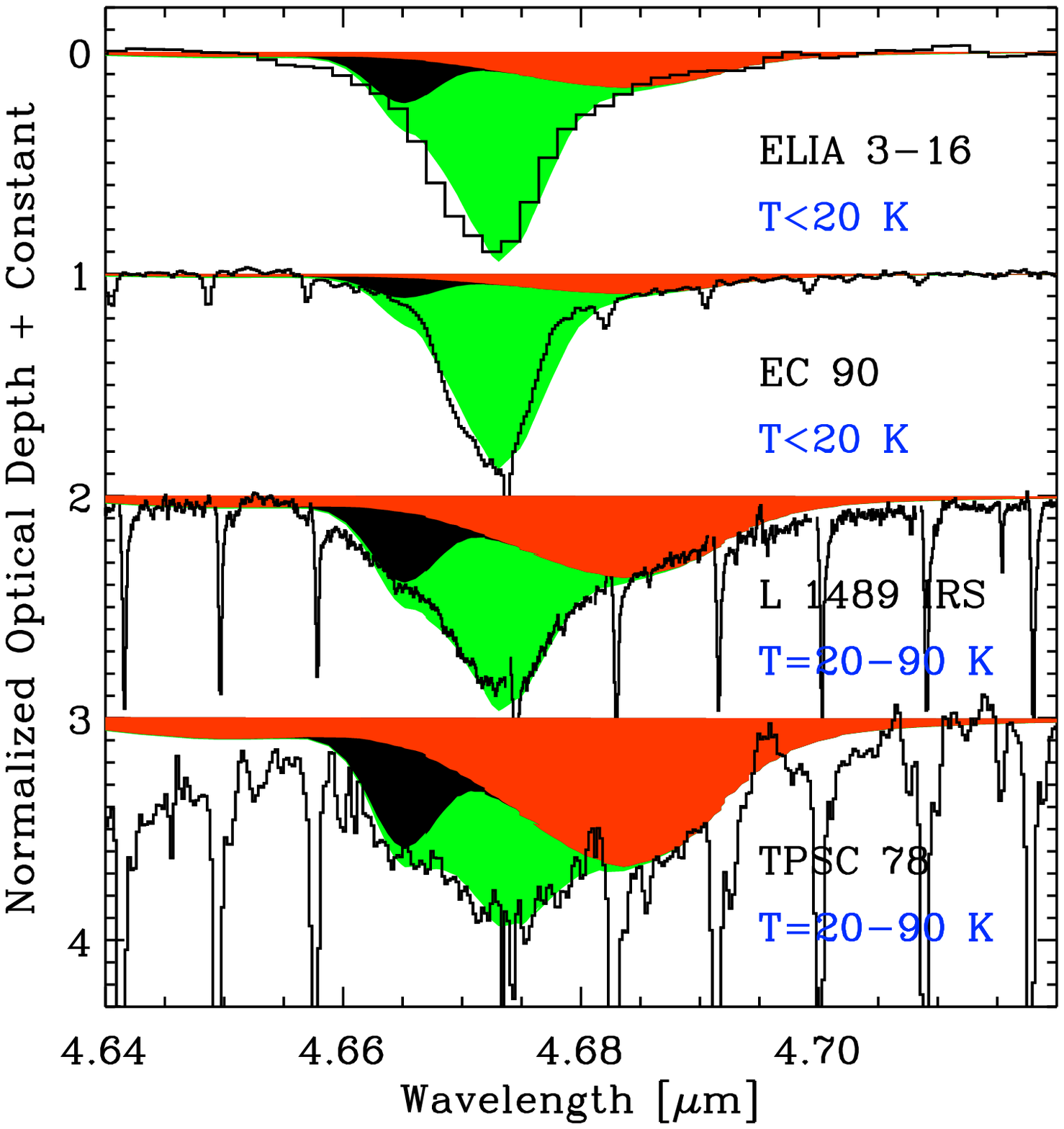,height=21pc,angle=90}}
\caption{The CO ice band profile varies strongly with line of sight
  temperature. The Taurus background star Elias 16 \citep{chi95} and
  the LYSO EC 90 \citep{pon03} trace dust below the pure CO
  sublimation temperature of $\sim 20$ K, while the LYSO L 1489 IRS
  \citep{boo02b} and the intermediate mass YSO TPSC 78 \citep{pon03}
  have temperatures primarily above that.  Two laboratory spectra are
  used to fit the profiles: pure CO corrected for CDE grain shapes
  (green) and CO in polar ices (red).  The polar ice is H$_2$O:CO=4:1
  in this case, but mixtures with CH$_3$OH are likely more appropriate
  (\S\ref{sec:467}). The black-colored feature on the short-wavelength
  side is a Gaussian representing a CO ice that probably contains
  CO$_2$, for lack of appropriate laboratory spectra. Ice sublimation
  is also evident by the increasing depth of the narrow ro-vibrational
  lines from gas phase CO.}\label{f:co}
\end{figure}

The 4.67 \mum\ band consists of three components as a result of solid
CO in different molecular environments. Gaussian representations of
these components fit all observed CO ice bands toward YSOs and
background stars (Table~\ref{t:feat}; Fig.~\ref{f:co}), and their
relative strengths may be used to derive the ice composition and dust
temperature along the line of sight \citep{pon03}. 
\marginpar{\textcolor{blue}{\bf Polar ices:} ices dominated by molecules with high
  dipole moments (e.g., H$_2$O, CH$_3$OH), which have relatively high
  sublimation temperatures.}
A narrow feature
centered on 4.673 \mum\ and a much broader long wavelength wing
peaking at 4.681 \mum\ are traditionally attributed to CO embedded in
low (apolar) and high (polar) dipole moment species \citep{san88}. The
sublimation temperature of the polar component is $\sim$90 K, much
higher than that of the apolar component ($\sim$20 K), and thus the
ratio of their column densities is often used to trace the ice
temperature (Fig.~\ref{f:co}; \citealt{tie91, chi98}).  
For the polar
CO component, entrapment or migration of CO into the abundantly
present H$_2$O ice seems an evident explanation, but the expected band
at 4.647 \mum\ (2152 \waven) due to dangling OH bonds in interstellar
spectra is missing \citep{san88}. Mixtures with CH$_3$OH-rich ices are
more likely, both spectroscopically and chemically
(\S\ref{sec:modelacc}; \citealt{cup11}).  High spectral resolution
observations have resolved the apolar component into a blue-shifted
shoulder at 4.665 \mum\ and a very narrow feature at 4.673
\mum\ \citep{boo02b}. The latter traces nearly ($\geq$90\%) pure CO.
\marginpar{\textcolor{blue}{\bf Apolar ices:} ices dominated by
  molecules with low dipole moments (e.g., CO, N$_2$), which have
  relatively low sublimation temperatures.}  The detection of the
solid $^{13}$CO band at 4.78 \mum\ was pivotal in this assessment
\citep{boo02a}. The absorption profile of this trace species is
insensitive to particle shape effects, and very sensitive to the ice
composition.  For the $^{12}$CO band, grains shaped according to the
Continuous Distribution of Ellipsoids (CDE) in the Rayleigh limit
provide the best fit.  The composition of the 4.665 \mum\ component is
less clear. It most likely contains CO$_2$, supported by a good
correlation of its strength with that of the CO:CO$_2$ component in
the 15.2 \mum\ CO$_2$ bending mode \citep{pon08}, but a comparison
with laboratory spectra shows that additional species are required in
the mixture as well \citep{boo02b}.  An alternative explanation by
absorption of linearly polarized light in a crystalline CO ice phase
(the longitudinal optical mode; \citealt{pon03}) appears less likely.

\subsection{The 15.2 \mum\ CO$_2$ Band: Segregation and Distillation}\label{sec:15}

\begin{figure}[b!]
\centerline{\psfig{figure=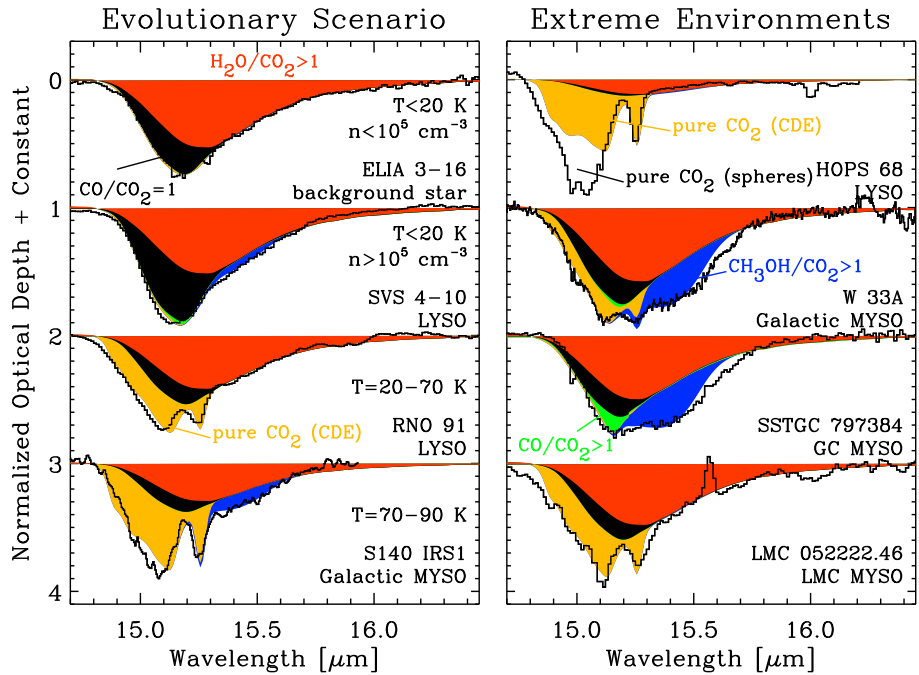,height=21pc,angle=90}}
\caption{CO$_2$ bending mode profiles observed in a range of
  environments. The colored areas show the components
  (\S\ref{sec:15}): polar (red), apolar (black), dilute (green), pure
  (yellow), and shoulder (blue). The left panel highlights an
  evolutionary scenario, showing, from top to bottom, the effects of
  increasing density and subsequent heating. The right panel shows
  sources that have experienced an extreme heating event (HOPS 68),
  have a high CH$_3$OH abundance (W 33A), or high (SST GC 797384) and
  low (LMC 052222.46) metallicities. The data were taken from
  \citet{ger99}, \citet{ber05}, \citet{pon08}, \citet{an11},
  \citet{sea11}, and \citet{pot13}.}\label{f:15um}
\end{figure}

\marginpar{\textcolor{blue}{\bf Particle shape effects:} band profile
  changes caused by resonances between the molecular dipoles and the
  particle charge distribution induced by the radiation field
  \citep{boh83}.}

The 15.2 \mum\ band of CO$_2$ has proven to be a particularly powerful
tracer of the composition and processing history of the ices
(Fig.~\ref{f:15um}). Building on initial work on ISO observations of
MYSOs by \citet{ger99}, \citet{pon08} use the much larger sample
observed by {\it Spitzer} toward LYSOs for an empirical decomposition
of the complex band profiles.  All observed bands can be fitted with a
combination of 5 laboratory spectra, that are fixed in position and
shape (Table~\ref{t:feat}), but whose relative abundances vary as a
result of the conditions along the line of sight (Fig.~\ref{f:15um}):
CO$_2$ (``pure''), CO$_2$/CO$\sim$1 (``apolar''), CO$_2$/CO$<<1$
(``dilute''), CO$_2$/H$_2$O$<1$ (``polar''), and CO$_2$/CH$_3$OH$<1$
(``shoulder'').  The dominance of the polar component is strongly
supported by the correlation of the CO$_2$ and H$_2$O column densities
(\S\ref{sec:abunvar}). A subset of YSOs shows distinct peaks at 15.11
and 15.26 \mum, which are attributed to those of pure CO$_2$ (the
symmetry of this linear molecule is broken in the pure ice
matrix). 
\marginpar{\textcolor{blue}{\bf Ice distillation:} sublimation of the most volatile
  species in mixtures (e.g., CO in CO$_2$ ice), leaving the species
  with higher sublimation temperatures behind.}  
These are the YSOs with the warmer envelopes, and it is thus
concluded that pure CO$_2$ ice is formed by the sublimation of CO from
apolar ices (``distillation''; 20 K) and at higher temperatures by
``segregation'' from polar ices as the H$_2$O crystallizes and the
rearrangement of bonds leads to clusters of pure CO$_2$
(\citealt{ehr99}; 30-77 K).  A few sources show a third peak at 15.35
\mum, which is attributed to bonding of the carbon atoms of CH$_3$OH
with the oxygen atoms of CO$_2$ (the ``shoulder'' component). 
\marginpar{\textcolor{blue}{\bf Ice segregation:} rearrangement of molecular bonds due
  to heating, leading to separation of species in ice mixtures,
  usually accompanied by crystallization.}
This is
not an indicator of thermal processing \citep{dar99}. Finally, the
``apolar'' component is probably responsible for the 4.66
\mum\ component of the CO band (\S\ref{sec:467}) as well, and indeed
their strengths correlate \citep{pon08}.

The profiles of the CO$_2$-rich components (``apolar'' and ``pure'')
of the strong 15.2 \mum\ band are affected by grain shapes in addition
to composition and thermal history.  As for CO, good fits are obtained
by CDE shaped grains in the Rayleigh limit.  The essential
confirmation comes from observations of the $^{13}$CO$_2$ band at 4.38
\mum\ \citep{boo00a}, which is insensitive to the grain shapes and can
be fitted with the same CO$_2$ components.  One striking exception of
the CDE model is the LYSO HOPS 68, which shows convincing evidence for
a population of spherical grains, reflecting an unusual history of
sublimation and rapid re-condensation (Fig.~\ref{f:15um};
\citealt{pot13}).

\marginpar{\textcolor{blue}{\bf ISO:} Infrared Space Observatory (1995-1998),
  operating spectrometers at wavelengths of 2.3-200 \mum\ and
  resolving powers of up to $R=\lambda/\Delta\lambda\sim$2000.}

\subsection{The 6.0 and 6.85 \mum\ Features: Complexity Revealed}\label{sec:6068}

\begin{figure}[t!]
\centerline{\psfig{figure=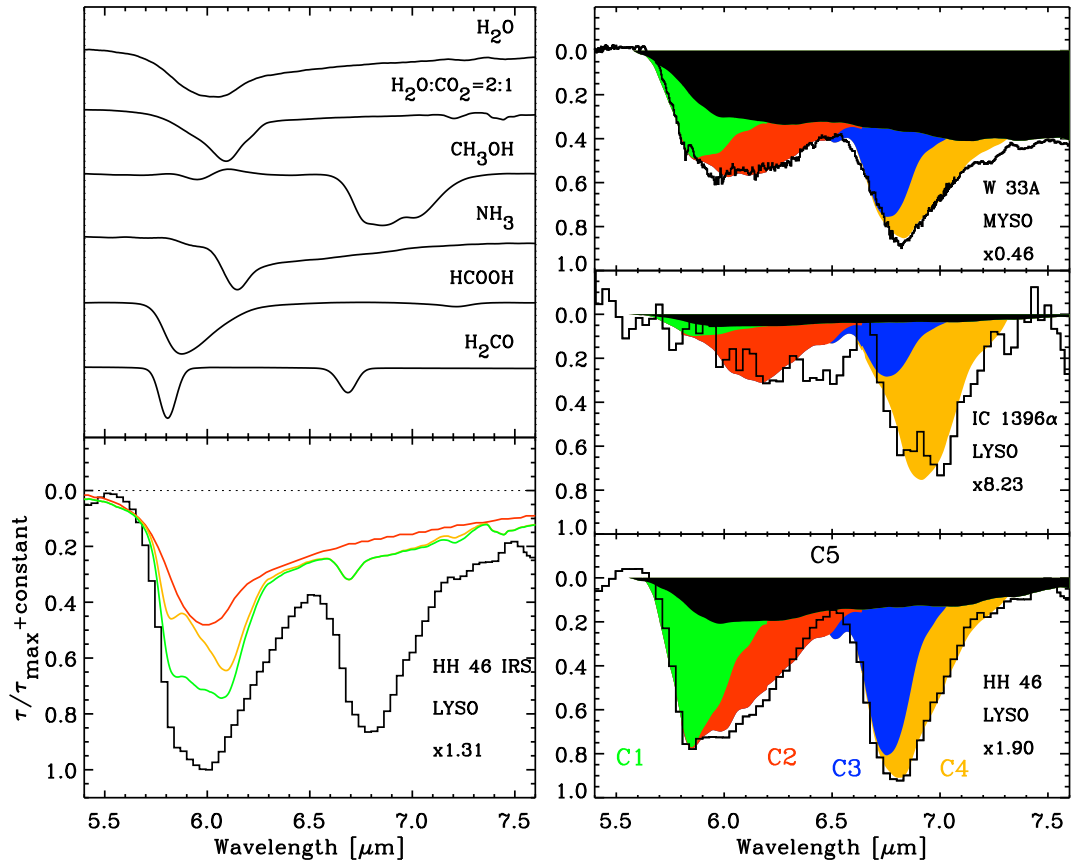,height=27pc,angle=90}}
\caption{{\bf Left Panels:} Laboratory spectra of species with modes
  in the 5-8 \mum\ region. The sum of these is compared to the
  observed 5-8 \mum\ features toward the LYSO HH 46 IRS \citep{boo08}.
  The pure H$_2$O contribution is given in red and the sum of all
  laboratory spectra including and excluding the uncertain HCOOH
  identification in green and yellow, respectively. A significant
  amount of absorption is unaccounted for. {\bf Right Panels:} Example
  of the decomposition of the H$_2$O-subtracted spectra of the MYSO W
  33A \citep{kea01b}, the LYSO IC 1396$\alpha$ \citep{rea09}, and the
  LYSO HH 46 in five empirically determined components C1-C5, used to
  investigate source-to-source variations of the band
  profiles.}\label{f:58um}
\end{figure}

The bending mode of H$_2$O peaks at 6.0 \mum\ and overlaps with an
H$_2$O combination mode that extends to 8 \mum. These are much weaker
than the 3.0 \mum\ stretch mode (\S\ref{sec:30}), because they benefit
less from the hydrogen bonding network \citep{thi57}. H$_2$O
absorption being less dominant, the 5-7 \mum\ region has proven to be
a good tracer of the complex composition of the ices
(Fig.~\ref{f:58um}), despite considerable identification issues
(\S\ref{sec:iden}).

High spectral resolution observations of MYSOs with ISO revealed
variations of the peak position of the 6.85 \mum\ feature
\citep{kea01b}, which could be attributed to the thermal
history. Also, a prominent broad absorption overlying both the 6.0 and
6.85 \mum\ features was discovered in a few sources \citep{gib02,
  gib04}.  
\marginpar{\textcolor{blue}{\bf Spitzer}: Spitzer Space Telescope, operating the IRS
  spectrometer (2003-2009) at $R\sim$60-128 (5-14 \mum) and $\sim
  100-600$ (10-37 \mum).}
The 5-7 \mum\ absorption observed toward LYSOs and
background stars with {\it Spitzer} was decomposed in five components,
after the subtraction of pure H$_2$O ice (Fig.~\ref{f:58um};
\citealt{boo08}), essentially confirming the earlier results toward
MYSOs: C1 (5.84 \mum) and C2 (6.18 \mum), C3 (6.755 \mum) and C4
(6.943 \mum) and component C5 peaking at 5.9 \mum\ but stretching over
the full 5.8-8 \mum\ range. Each component probably represents more
than one carrier.  The C1 and C3 components vary little relative to
the H$_2$O column density.  Component C4 and perhaps C2 become more
prominent at low H$_2$O ice abundances, i.e., at temperatures above
H$_2$O sublimation ($>90$ K). These features can thus be treated as
tracers of extreme conditions.  Such cases have proven to be rare for
LYSOs.  Enhancements of the C5 component do not appear to correlate
with those of C4, and may thus trace different processing conditions
(e.g., at lower temperatures or by energetic photons or
particles). Measurements of the strength and shape of this component
are inaccurate, however, because it is so wide and overlaps with the
H$_2$O bending mode (Fig.~\ref{f:58um}).

\subsection{The Lattice Modes: Ice Structure and Emission}\label{sec:lattice}

Although the lattice modes were detected early on (in Orion-KL;
\citealt{eri81}), relatively few observations were published due to
the limited availability of telescope instrumentation at wavelengths
above 25 \mum. These observations do show unequivocally the
sensitivity of the modes to the thermal history of the ice
\citep{mal99, dar98, omo90, syl99, dem00, chi01}.  The H$_2$O feature
at 62 \mum\ is particularly strong in crystalline ices.  Laboratory
experiments have shown that the bands are very sensitive to the ice
composition as well, because impurities affect the long range modes
\citep{moo94, iop14}. In fact, along with the torsional modes at
similar wavelengths, they identify complex species more uniquely than
do the mid-infrared intra-molecular vibrations.  Finally, as the
lattice modes may appear in emission, they show great promise as tools
to map the ice properties.

\section{ICES ACROSS ENVIRONMENTS}~\label{sec:env}

Ices are present in a wide range of environments outside of the solar
system: envelopes and disks of young and evolved stars and Galactic
and extragalactic molecular clouds and cores (Fig.~\ref{f:3um}).
These environments are characterized by high densities ($\geq 10^3$
\cubcm), low temperatures ($<90$ K) and weak UV radiation fields
($G_{\rm 0}\leq 0.07$ at \av$\geq 1.6$ mag; \S\ref{sec:modelen};
\citealt{hol09}). The same species are generally detected in all
environments, with the exception of the envelopes of evolved stars
which only show H$_2$O ice.  The ice abundances vary greatly, however,
both across and within environment classes.  This reflects differences
in physical conditions, age, and elemental abundances. Observational
studies of the different environments are briefly reviewed here, and
abundances are summarized in \S\ref{sec:abun}.

\subsection{Quiescent Dense Clouds and Cores}~\label{sec:bg}

\begin{figure}[t!]
\centerline{\psfig{figure=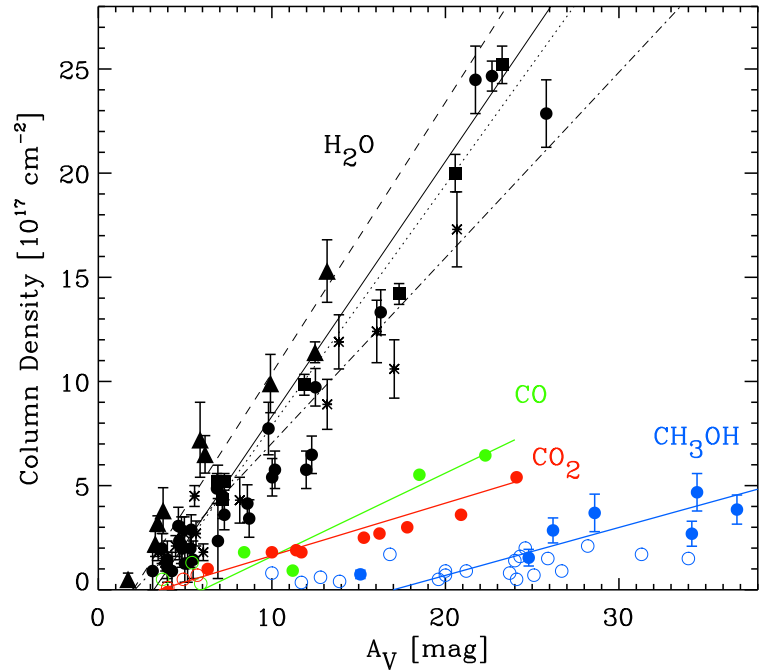,height=21pc,angle=90}}
\caption{Relation between H$_2$O (black), CO (green), CO$_2$ (red),
  and CH$_3$OH (blue) ice column densities and extinction \av\ for
  quiescent lines of sight in nearby clouds and cores. Upper limits
  are indicated by open symbols. The lines represent linear fits to
  the detections. For H$_2$O, data points are shown for the Taurus
  (circles and solid line) and Lupus IV (asterisks and dash-dot line)
  clouds, as well as the L183 (triangles and dashed line) and IC 5146
  (squares and dotted line) cores. For CO and CO$_2$, the data points
  are for Taurus only, and for CH$_3$OH they are for a variety of
  dense cores. The data were taken from \citet{chi95}, \citet{whi07},
  \citet{boo11}, \citet{chi11}, \citet{whi13} and
  \citet{boo13}.}\label{f:thresh}
\end{figure}

The first detections of the 3.0 \mum\ H$_2$O ice band toward
background stars showed that a dense cloud or core environment and a
threshold extinction are needed for the formation of detectable
quantities of H$_2$O ices \citep{whi83}.  In Taurus, the ice band is
only present at extinctions above \av$=3.2\pm0.1$ mag \citep{whi01},
which corresponds to an ice formation threshold of \av$=1.6$ mag
considering that background stars trace both sides of the cloud.
\marginpar{\textcolor{blue}{\bf Ice formation threshold:} cloud extinction depth
  ($A_{\rm V}$) indicating the onset of rapid ice mantle growth, after
  a monolayer has formed.}  Similar values are seen
(Fig.~\ref{f:thresh}) in the Lupus cloud \citep{boo13} as well as the
dense cores L183 \citep{whi13} and IC5146 \citep{chi11}.  Larger
thresholds claimed for some environments are not as well established.
Toward the Oph cloud, generally considered a high radiation
environment hindering ice growth, few background stars were observed
and threshold determinations include YSOs \citep{tan90}.

Solid CO$_2$, which was found to be ubiquitous in quiescent dense
clouds \citep{whi98}, has effectively the same formation threshold as
H$_2$O \citep{ber05, whi09}, showing a close chemical connection
between these species.  
\marginpar{\textcolor{blue}{\bf Monolayer:} one layer of molecules covering dust grain
  surfaces, for H$_2$O corresponding to $10^{15}$ molecules/cm$^2$ and
  an abundance $X_{\rm H}\sim$few$\times 10^{-6}$.}
Still, their relative abundances vary by up to
a factor of three (\S\ref{sec:abunvar}). Freeze-out of the more
volatile CO species occurs deeper into the cloud, at $\sim 3$ mag
(Taurus, Serpens; Fig.~\ref{f:thresh}; \citealt{chi95}).
\citet{ker93} determine a three times larger threshold in Ophiuchus,
but along individual lines of sight, values are lower, probably due to
clumpiness \citep{shu00}.  CH$_3$OH ice forms at even greater cloud
depths ($9\pm3$ mag; \citealt{boo11, chi11, whi11b}), consistent with
the formation requirement of ``catastrophic'' CO freeze out
(\S\ref{sec:modelacc}).  This is reflected in large CH$_3$OH abundance
variations, from less than 3\% in Taurus \citep{chi96} to 12\% in some
isolated dense cores \citep{boo11, chi11}.

A general characteristic is the pristine nature of the ices in these
environments: the dominance of amorphous H$_2$O \citep{smi93}, the
large contribution of apolar ices to the CO band \citep{whi85}, and
the absence of segregated (heated) CO$_2$ ices \citep{whi09}.  The
peak position of the 6.85 \mum\ feature is also consistent with
unprocessed ices \citep{kne05, boo11}. The absence of the 4.62
\mum\ ``XCN'' feature \citep{whi85, nob13} is sometimes interpreted as
a lack of processing, but, more likely, it is related to insufficient
CO freeze-out (\S\ref{sec:iden} and \S\ref{sec:model}).

\subsection{Envelopes of Massive YSOs in the Galactic Disk}~\label{sec:mysos}

A sample of $\sim$20 MYSOs (luminosities above $10^4$ L$_\odot$,
masses above 10 M$_\odot$) in the Galactic Disk has been studied
extensively in the ice features, using mostly data from ISO. We refer
to reviews by \citet{gib04}, \citet{boo04b}, and \citet{dar05} for
historical work.  Few new observations of these objects have been
published since the launch of Spitzer Space Telescope in 2003, but the
earlier data have been re-analyzed with new methods or included as
comparison objects with LYSOs in later work \citep{pon08,
  boo08}. MYSOs in the GC region are discussed in \S\ref{sec:gc}.

The evidence for thermal processing of the MYSO ices is ubiquitous in
all the tracers discussed in \S\ref{sec:prof}. This is mostly based on
observations along a single line of sight toward the central object,
but ice maps have confirmed it.  Using extended scattered light as a
background source, the 3.0 \mum\ band is deep toward a disk-like
structure around the MYSO AFGL 2136, but the ice abundance is
suppressed relative to the surrounding cloud \citep{hol98}, probably
as a result of sublimation.  Similar conclusions were drawn for ice
maps toward the MYSO S140 IRS1 \citep{har97}.

\subsection{Envelopes of Low and Intermediate Mass YSOs}~\label{sec:lysos}

As the link between quiescent clouds and protoplanetary disks that
give birth to potentially life-bearing planetary systems, ices in the
envelopes of LYSOs are of paramount importance. The first (3.0 \mum)
ice band toward an LYSO was detected by \citet{coh75} toward HL Tau
and it was thought to originate in an edge-on disk \citep{coh83}, but
a location in the envelope is more likely \citep{fur08}. This
demonstrates the difficulty of locating ices along the line of sight
\citep{boo02c}. Here, surveys of ices most likely located in envelopes
around LYSOs are reviewed, while disk studies are reviewed in
\S\ref{sec:disks}.

Comprehensive surveys of ices in the envelopes of embedded LYSOs were
enabled by the {\it Spitzer}  mission (5-20 \mum) and sensitive ground-based
3-5 \mum\ instruments.  The combination of the observations proved
critical in the separation of overlapping bands (e.g., in the 5-8
\mum\ region; \S\ref{sec:iden}), in obtaining stronger constraints on
the ice mantle composition (e.g., unraveling the 15.2 \mum\ CO$_2$
band profile; \S\ref{sec:15}), and in obtaining a complete
inventory. Samples of $\sim 45$ objects in nearby clouds (Perseus,
Serpens, Corona Australis, Taurus, Oph) and several cores were studied
in the 5-8 \mum\ and H$_2$O features \citep{boo08} and the bands of CO
and CO$_2$ \citep{pon03, pon08}, CH$_4$ \citep{obe08}, CH$_3$OH and
NH$_3$ \citep{bot10}, and ``XCN'' \citep{bro05}. A survey of all ice
absorptions in the 5-20 \mum\ range focused on LYSOs in Taurus was
done as well \citep{zas09}.

These studies consider mostly Class 0 and I LYSOs. As for MYSOs,
variations in abundances and processing history were observed.
Thermal processing was found to play an important role, despite the
lower stellar luminosities. Aided by episodic heating events,
sublimation of the most volatile CO and CO$_2$ components was observed
for even the lowest luminosity YSOs ($<$0.1 L$_\odot$;
\citealt{kim12}).  Large variations of the OCN$^-$ abundance (4.64
\mum), even for sources close together on the sky, suggest special
conditions in envelopes \citep{bro05}. They may be related to past CO
freeze-out and possibly heating events (\S\ref{sec:model}).

LYSOs in clustered and high radiation environments of nearby massive
stars may be more representative of the early solar system than
isolated YSOs \citep{ada10}. An example is the IC~1396 globule
irradiated by a nearby O-star. It shows 6.85 \mum\ features peaking at
relatively long wavelengths \citep{rea09}, a high temperature effect
that is rarely observed toward isolated LYSOs (\S\ref{sec:6068}).

Finally, observations of the envelopes of intermediate mass YSOs
(masses of 2-10 M$_\odot$ and luminosities of a few to $10^4$
L$_\odot$) would enable further studies of the dependency of ice
characteristics on stellar mass, but the sample sizes are small.  In a
{\it Spitzer}  survey of 14 embedded intermediate mass YSOs distributed on
the sky, \citet{pit11} showed that the common ice features are present
and that the 15.2 \mum\ CO$_2$ band shows no signs of segregation.
Thermal processing is, however, evident by reduced H$_2$O and CO ice
abundances in intermediate mass YSOs in the Vela cloud
\citep{thi06}. One of these sources has high CH$_3$OH and OCN$^-$
abundances, but this is not necessarily a sign of thermal processing
(\S\ref{sec:iden} and \ref{sec:modelacc}).

\subsection{Disks Around Low and Intermediate Mass YSOs}~\label{sec:disks}

Direct observations of ices in circumstellar disks trace the volatile
building blocks of planet and comet formation.  These observations are
challenging because the relatively flat disk geometry requires a
specific orientation for the 2-20 \mum\ vibrational modes to be seen
in absorption. Great improvements in the sensitivity and spatial
resolution of infrared observations, as well as in disk models, has
led to much progress, however. A recurring conclusion is that ices in
disks are strongly affected by thermally processing.

For flared disks, the observed ice features are most prominent if the
disk inclination is comparable to the opening angle of about 70
degrees \citep{pon05}. Lower angles trace less dense outer disk
regions and the observed ice features are also diluted as the size of
background scattered light region exceeds that of the absorbing
region. At higher angles, the high disk mid-plane densities increase
the dust optical depths. As a result, ice observations are biased
toward processed ices in the disk surface layers. This is indeed
observed in the edge-on disk source CRBR 2422.8-3423 in Oph
\citep{pon05}, as it is one of the few low mass sources with a shifted
6.85 \mum\ feature indicative of high temperature processing
(\S\ref{sec:6068}).

Rather than using the light from the central object, \citet{ter12a}
and \citet{ter12b} use the extended background of the Orion nebula to
trace the ices. This work confirms the optimal disk inclination angle
of 65-75 degrees needed to detect ice features. One source shows an
unprecedented H$_2$O ice band profile (Fig.~\ref{f:3um}) with a strong
contribution from large ($\sim 0.8$ \mum) grains rich in crystalline
ices that may have surfaced from the lower disk regions where
coagulation is more efficient. The same authors have resolved a
(different) edge-on disk against the extended background and find
spatial variations in band depth but not shape.  The depth does not
vary on a time scale of 4 years \citep{ter12b}, implying that the
crystalline ices are extended over at least 3 AU at distances beyond
30 AU.  This distance estimate comes from spatially resolved
spectroscopy by \citet{sch10} toward the T Tau object YLW 16A.

High spatial resolution observations have indeed been very powerful in
locating ices in disks. Coronographic adaptive optics (AO) images in
narrow band filters of the Herbig Ae star HD 142527 reveal a spatially
extended 3 \mum\ absorption attributed to scattering by icy $\sim 1$
\mum\ sized grains in the disk \citep{hon09}. Also, AO spectra
separated the HK Tau and HV Tau binary and tertiary systems, revealing
deep (amorphous) ice bands in the HK Tau B and HV Tau C edge-on
disks. The depth toward HV Tau C varies on a time scale of 2.3 years,
which could be caused by a 1.4 AU sized structure at a distance of 100
AU from the star \citep{ter07}.

A particularly successful technique to trace ices in circumstellar
disks, that is less dependent on the disk orientation, is spectroscopy
of the far-infrared ice lattice modes (\S\ref{sec:lattice}). In
passively heated, flared disks, the lattice modes of H$_2$O ice at 44
and 62 \mum\ are excited in super-heated optically thin disk surfaces
at radii beyond 100 AU \citep{chi01}. In this layer, the ice
condensation radius (not to be confused with the ``snow line'' in the
disk mid-plane) scales with the 3rd power of the stellar temperature,
which explains the detection of the lattice modes in ISO spectra of
the coolest members of their sample of Herbig AeBe and T Tauri stars.

The ISO spectrometers detected more of the lattice modes, mostly from
crystalline H$_2$O, in bright disks around intermediate mass stars:
the isolated Herbig Ae stars HD 100546 \citep{mal98} and HD 142527
\citep{mal99}, and the Herbig Ae star HD 163296, but not AB Aur
\citep{anc00}.  Observations with the Herschel/PACS spectrometer are
limited by calibration uncertainties and incomplete coverage of the
strongest (44 \mum) lattice mode. Detections claimed for a few T Tauri
stars indicate a high degree of crystallinity created by past heating
events, such as planetesimal collisions \citep{mcc15}.

Finally, sensitive, but low spatial resolution {\it Akari}
observations of a sample of edge-on disk sources confirm the optimal
inclination that is needed to detect ice features, but the spectral
resolution is too low to study processing effects \citep{aik12}.
\marginpar{\textcolor{blue}{\bf Akari}: a Japanese space telescope, operating the
  InfraRed Camera (2006-2011) equipped with, among others, a prism at
  $R\sim$20-100 (2.5-5 \mum).}
Their 2-5 \mum\ spectra reveal solid H$_2$O, CO, OCN$^-$, OCS, and
possibly HDO.

\subsection{Galactic Center Region}~\label{sec:gc}

The conditions in the GC region differ significantly from those
elsewhere in the Galaxy (elemental abundances, radiation fields,
dynamic time scales) and thus offer a unique perspective on ice
formation and evolution processes.  Studies of ice features in the GC
region have primarily targeted the cluster of infrared sources (GC
IRS) within a parsec from SgrA*, and MYSOs (candidates) throughout an
extended ($\sim 50\times $200 pc) region in the Central Molecular Zone
(CMZ).

Most of the infrared sources in the central parsec of the Galaxy show
ice absorption features. The depths of the 3.0 \mum\ H$_2$O and 4.67
\mum\ CO bands \citep{mcf89, chi02, mou05} and the 4.62 \mum\ XCN
feature \citep{mon01, chi02, mou09} vary with factors of $\sim$4
between these sources. As much as 60-80\% of the H$_2$O ices may be
local to the inner pc of the Galaxy, possibly in dense clumps in the
circum-nuclear ring or mini-spiral \citep{chi02, mou05}. The
foreground contamination is less certain for CO ices \citep{mou09},
but in any case the total column relative to H$_2$O is small
($<10$\%). While not much CO has frozen out, there is little evidence
for thermal processing. The CO profile has a strong apolar component
\citep{mon01, mou05}, and the 15.2 \mum\ CO$_2$ bending mode, observed
in the ISO beam of $\sim 20$'' encompassing a number of IRS sources,
shows a lack of thermal processing as well \citep{ger99}. Other
detected ice species include NH$_3$ and CH$_4$, while solid CH$_3$OH
is notably under-abundant ($<5$\%; \citealt{chi00, chi02}).  The
detection of the stretch mode of NH$_3$ at 2.96 \mum\ in ISO and high
spatial resolution ground-based observations, supported by the
presence of NH$_3$.H$_2$O hydrates at 3.2-3.7 \mum\ and by the 6.2
\mum\ bend mode, is unusual and implies a large NH$_3$ abundance of up
to 30\% relative to solid H$_2$O. An N-rich environment seems also
reflected in a high OCN$^-$ abundance of 3\% toward GC IRS 19
(\citealt{chi02}, recalculated using the band strength in
\citealt{bro05}).

On a larger scale, ices were detected in numerous sources throughout
the CMZ. In a sample of 107 targets selected using {\it 2MASS} and
{\it Spitzer} 1-8 \mum\ broad band photometry, 35 were identified as
MYSOs (or candidates; 8-23 $M_\odot$) following {\it Spitzer}
spectroscopy \citep{an11}. They are located throughout the CMZ,
although there is clustering near the Sgr B molecular complex. All
objects show ice features in the 5-7 \mum\ region, as well as the 15.2
\mum\ CO$_2$ bending mode. The latter was used as a criterion in the
classification of the sources as MYSOs: the presence of a long
wavelength wing tracing the CH$_3$OH:CO$_2$ complex was required
(Fig.~\ref{f:15um}). It is based on the fact that until then this wing
had only been detected towards Galactic Disk MYSOs with high CH$_3$OH
abundances (\S\ref{sec:15}). Further work is needed to confirm the
nature of these sources and the location of the ices in the MYSO
envelopes or in the dense cloud medium in the CMZ.  Also, spectra at
wavelengths shorter than 5 \mum\ have not yet been obtained, thus
direct measurements of the H$_2$O, CH$_3$OH, and CO abundances toward
the CMZ MYSOs are lacking.

\subsection{Shells Surrounding Evolved Stars}~\label{sec:evolved}

Certain classes of evolved stars show H$_2$O ice bands and offer a
view on ice properties from a completely different perspective
compared to that of the ISM and YSOs. In general, these are
carbon-poor AGB stars with very high mass loss rates, showing OH maser
emission (OH/IR stars), or the slightly more evolved post-AGB stars
that still have dense, but expanded and cooled envelopes.  In short,
the expanding dense shells of gas expelled by evolved stars cool
rapidly. At some point, H$_2$O is formed in the warm gas, and as the
shell cools it condenses onto the refractory grains formed in an
earlier, hotter phase. The 3.0 \mum\ band is narrower in OH/IR stars
compared to YSOs (Fig.~\ref{f:3um}; \citealt{gil76}) and is thought to
reflect the purity of the H$_2$O ice, its crystallinity, and the small
grain size \citep{smi88}.  The 6.0 bending mode \citep{soi81}, the 11
\mum\ libration mode \citep{syl99} and the lattice modes at 44 and 62
\mum\ were detected as well.  The lattice mode was first detected by
anomalously strong IRAS 60 \mum\ broad band emission, half of which
can be attributed to H$_2$O ice \citep{for87, omo90}, in the post-AGB
star Frosty Leo.  The lattice modes strongly confirm the crystallinity
of the ice \citep{omo90, syl99, dem00}. No other ice species were
detected toward evolved stars \citep{gil76, syl99}, in part due to the
oxygen rich environment in which they were formed.

\subsection{Massive YSOs in the Magellanic Clouds}~\label{sec:mc}

Observations of ices in the Magellanic Clouds offer unique tests of
ice and molecule formation and evolution models, due to the lower
metallicities and correspondingly lower dust content and higher UV
fields. Overall, differences in the relative ice abundances and higher
levels of processing have been found relative to Galactic ices.

The CO$_2$/H$_2$O column density ratios are a factor of two larger in
the envelopes of MYSOs ($10^4-10^5~L_\odot$) in the Large Magellanic
Cloud (LMC; 33$\pm$1\%; \citealt{sea11} using data from
\citealt{shi08, oli09, shi10}) compared to the Galaxy (17$\pm$3\%;
\citealt{ger99}). The shape of the 15.2 \mum\ CO$_2$ band is very well
studied (\S\ref{sec:abunvar}; \citealt{sea11, oli09}).  In the
majority of sightlines, CO$_2$ ices have experienced thermal
processing.  CO ice is detected in ground-based spectra \citep{oli11}
and in low resolution {\it Akari} spectra \citep{shi10}, indicating a
low abundance ($\leq 15$\% relative to H$_2$O). The 5-7 \mum\ ice
features, including a clear 6.85 \mum\ feature in at least one object,
were detected \citep{oli09} although their analysis is severely
hampered by overlapping PAH emission \citep{sea09}. The XCN feature at
4.62 \mum\ may have been detected in a few {\it Akari} sources but
suffers from low spectral resolution \citep{shi10}.  This is the case
for the 3.53 \mum\ band of solid CH$_3$OH as well.

H$_2$O and CO$_2$ ices were also detected toward 14 MYSOs in the Small
Magellanic Cloud (SMC) in a combined {\it Spitzer} and ground-based
3-4 \mum\ survey \citep{oli13}. The relative column densities are
compared to other environments in \S\ref{sec:abunvar}.  Heating
effects play a role here as well, judging the presence of the 62
\mum\ lattice mode of crystalline H$_2$O ice in {\it Spitzer} spectra
\citep{loo10}.  The 6.0 and 6.8 \mum\ features were detected, but CO
ice is absent at levels below those in the LMC (although only four
lines of sight were observed; \citealt{oli11}).

\subsection{Other Galaxies}~\label{sec:extgal}

Ice features are commonly seen in the infrared spectra of external
galaxies with centers dominated by starbursts or dust-enshrouded
Active Galactic Nuclei (AGN). They are often used to determine the
nature of the emission sources. \citet{ima03} show that peak optical
depths of the 3.0 \mum\ band larger than $\sim 0.3$ must be caused by
compact AGN emission surrounded by dust clouds containing the
ices. Weaker ice features are due to starburst regions in which many
MYSOs are mixed with clouds containing ice coated dust grains within
the same beam.  Local ULIRGS and most LINER-type galaxies show deep
ice bands, and their centers are thus dust-enshrouded AGN. In
contrast, Seyfert-type galaxies do not show ice bands at all, lacking
such dusty centers \citep{ima06}.

Besides H$_2$O ice (3.0 and 6.0 \mum), solid CO$_2$, ``XCN'', CO and
CH$_4$ were detected in these galaxies \citep{spo00, spo04}.  The
spectra often show a combination of PAH emission, and absorptions from
diffuse dust and ices, much like the center of the Galaxy
\citep{stu00}, complicating the analysis of the ice features.

The ice distribution was mapped in several nearby galaxies, and
enhanced relative abundances are often attributed to production in
energetic radiation fields.  Toward a ring encircling the starburst
nucleus of NGC 4945, the 4.62 \mum\ XCN feature is stronger than that
of CO, similar to that in the Galactic MYSO W33A. The depths do not
vary much within the ring. \citet{spo03} attribute it to UV-processed
ices (X-ray fields are likely too weak). The {\it Akari} satellite
mapped the ice distributions in other starburst-dominated galaxies. In
M82, variations of the CO$_2$/H$_2$O abundance ratio correlate with
the UV field traced by Br$\alpha$ emission, possibly indicating the
energetic production of CO$_2$ \citep{yam13}.  In view of the
difficulty of identifying energetically-produced ices in Galactic
environments (\S\ref{sec:modelen}), these claims for nearby galaxies
must be regarded as tentative.

The current redshift record for ice feature detections is 2.28
\citep{saj09}. At such high redshift, both the 3 and 6 \mum\ ice bands
are within the {\it Spitzer} spectrometer wavelength range. Overall,
no differences in the ice composition or abundance were found as a
function of redshift, but the data are limited.

\subsection{Solar System Objects}~\label{sec:ss}

There are many ice-rich environments in the Solar System, including
the satellites of the gas- and ice-giant planets, trans-Neptunian
Objects (TNOs), and comets.  For this review of interstellar and
circumstellar ices, cometary ices are of particular interest because
of their possible link with ices in the ISM.  Ice abundances are the
main observables to test this link.  While present day comets
originate from distinct heliocentric distances, formation regions are
less clear as a result of radial mixing. For example, organic-enriched
comets are present in both Kuiper Belt and Oort Cloud comets.  A
certain population of comets may even be captured from sibling star
systems formed from the same cluster as the Sun. Cometary ice
abundances will not be reviewed here, but rather the review by
\citet{mum11} will be used as a reference. Remarkable differences and
similarities with respect to interstellar ices are observed
(\S\ref{sec:abunss}).

\section{COLUMN DENSITIES AND ABUNDANCES}~\label{sec:abun}

Ice column densities are generally determined by dividing the
integrated optical depth by the integrated band strength {\it A}
measured in laboratory experiments (\S\ref{sec:lab}):

\begin{equation}
N=\int\tau_\nu d\nu/A\ [{\rm cm^{-2}}]
\label{eq:colden}
\end{equation}

For saturated bands, and in particular for unresolved ones such as
those observed with the low resolution mode of the {\it Akari}
satellite ($R\sim 20$), a curve of growth analysis must be performed
to derive the column density \citep{shi08}. For LYSOs with disks, the
size of the background region of scattered light, and thus the
observed ice column in the foreground, may be a strong function of
wavelength \citep{pon05}. Different modes of the same species would
then yield different column densities, and in fact their ratio would
trace the disk inclination. So far this has not been observationally
demonstrated, however. Similar scenarios were suggested to explain
discrepancies between the 3 and 6 \mum\ absorption band depths towards
MYSOs due to H$_2$O \citep{kea01b}, but this also still remains to be
proven. For example, toward the MYSO S140 IRS1, the column densities
derived for all CO$_2$ bands over the 2-15 \mum\ wavelength range are
in good agreement \citep{kea01a}. Finally, for external galaxies such
as ULIRGs and starburst galaxies, a single spatial resolution element
may contain multiple continuum sources (e.g., YSOs, an AGN), and
depending on the geometry of those and the dense clouds containing
ices, the ice band depths may be quite different (\S\ref{sec:30};
\citealt{ima03}).

Equation~\ref{eq:colden} only applies to pure absorption.  For
features caused by scattering (\S\ref{sec:em}), the ice column traced
depends strongly on the grain size, and radiative transfer involving
grain size distributions and appropriate optical constants must be
calculated \citep{pen90, dar01}.  \marginpar{\textcolor{blue}{\bf
    MRN:} number density distribution of grain radii derived from
  diffuse cloud observations, n(a) $\propto a^{-3.5}$ \citep{mat77}.}
Similarly, the lattice modes ($>25$ \mum) may appear in emission and
radiative transfer models are required to evaluate column densities
such as demonstrated for disks surrounding YSOs \citep{chi01} and
envelopes surrounding MYSOs \citep{dar98}.

H$_2$O ice was detected at column densities ranging from $\sim
10^{17}$ to $10^{19}$ \sqcm. The lower limit corresponds to a peak
optical depth of the 3.0 \mum\ band of $0.06$, which is rather high
because of uncertainties in the continuum determination. It
corresponds to dust extinctions close to the ice formation threshold
(\S\ref{sec:bg}) and thus to a few monolayers of ice on each grain in
the MRN grain size distribution with a lower limit of 20
\AA\ \citep{hol09}. 
\marginpar{\textcolor{blue}{\bf Upper and lower quartile abundances:}
  median of the abundances respectively larger and less than the
  median of all abundances.}  H$_2$O might be present on grains below
the ice formation threshold, but, besides the low column, the
observational signature is weakened by the much lower band strength of
isolated water compared to that of bulk ice (\S\ref{sec:30}).

Ice abundances relative to H$_2$O, $X_{\rm H_2O}$, were determined
following the statistical method in \citet{obe11}, but using somewhat
different samples and including minor ice species.  The values for the
Galactic sources are listed in Table~\ref{t:abun}. For samples larger
than a few, the median and lower and upper quartiles of the abundance
distributions are given, as well as the median abundances taking into
account upper limits.  Also listed are the full ranges of the observed
abundances. 

Cometary abundances derived from measured gas phase
production rates were taken from \citet{mum11}, supplemented by
measurements of CO$_2$ abundances with the {\it Akari} satellite
(within 2.7 AU, the H$_2$O sublimation radius; \citealt{oot12}).
Considerable variations are observed within and across environments
(\S\ref{sec:abunvar}).  The insecurely identified species are
discussed in \S\ref{sec:iden}, and upper limits for other
astrochemically relevant species in \S\ref{sec:uplim}.

\begin{landscape}
{
\begin{longtable}{l|llll|lll}
\caption{Ice Abundances}\label{t:abun}\\
\toprule
Species & \multicolumn{4}{c|}{$X_{\rm H_2O}^{\rm a}$ [\%]}                                              & \multicolumn{3}{c}{$X_{\rm H}^{\rm b}$ [$10^{-6}$]}\\
        & MYSOs                  & LYSOs                      & BG Stars$^{\rm c}$  & Comets         & MYSOs                       & LYSOs                             & BG Stars$^{\rm c}$ \\        
\colrule
\multicolumn{8}{l}{{\bf Securely identified species$^{\rm d}$:}}\\
H$_2$O$^{\rm e}$ & 100             & 100                       & 100                & 100            & $31_{16}^{50}$               & $38_{25}^{55}$ (42)                 & $40_{31}^{48}$ (39)                 \\
            &                    &                           &                    &                & 12-57                       & 14-80                              & ($<9$)-62                             \\
CO$^{\rm e}$  & $7_4^{15}$  (7)     & $21_{12}^{35}$ (18)         & $25_{20}^{43}$       &               & $2.6_{0.6}^{8.8}$ (1.9)       & $9.6_{4.8}^{17}$ (8.1)              & $12_{9}^{20}$                        \\
        & 3-26                   & ($<$3)-85                 & 9-67               & 0.4-30         & ($<0.4$)-12.8               & ($<1.2$)-26                       & 3-21                               \\
CO$_2$$^{\rm e}$ & $19_{12}^{25}$    & $28_{23}^{37}$              & $26_{18}^{39}$       & $15_{10}^{24}$  & $3.7_{2.5}^{12}$              & $11.8_{6.0}^{20}$                    & $13.2_{8.2}^{18}$                    \\
        & 11-27                  & 12-50                     & 14-43              & 4-30           & 1.8-15.6                    & 2.4-25                            & 5.2-26                            \\
CH$_3$OH   & $9_5^{23}$ (5)       & $6_5^{12}$ (5)              & $8_6^{10}$ (6)      &                & $3.7_{1.9}^{11}$ (1.7)        & $3.3_{2.7}^{7.9}$ (2.3)             & $5.2_{3.2}^{6.4}$ (2.4)             \\
        & ($<3$)-31              & ($<1$)-25                 & ($<$1)-12          &  0.2-7         & ($<0.4$)-16.6               & ($<0.2$)-15                      & ($<0.6$)-6.6                      \\
NH$_3$  &                        & $6_4^8$ (4)                &                    &                &                             & $3.6_{2.4}^{5.4}$ (2.6)             &                                    \\
        & $\sim $7$^{\rm f}$      & 3-10                       & $<7$               &  0.2-1.4       & $\sim $4$^{\rm f}$           & ($<0.4$)-6.4                      & $<4$                               \\
CH$_4$     &                     & $4.5_3^{6}$ (3)            &                    &                &                             & $2.3_{1.5}^{3.7}$ (1.4)             &                                    \\
        & 1-3                    & 1-11                      & $<3$               &  0.4-1.6       & 0.4-1.8                     & ($<0.2$)-5.6                      & $<$1.2                           \\
\colrule
\multicolumn{8}{l}{{\bf Likely identified species$^{\rm g}$:}}\\
H$_2$CO & $\sim$2-7               & $\sim$6                   &                    & 0.11-1.0      & $\sim$2                    & $\sim$3                            &                           \\
OCN$^-$ & $0.6_{0.3}^{0.7}$         & $0.6_{0.4}^{0.8}$ (0.4)     &                    &               & $0.4_{0.2}^{0.5}$             & $0.4_{0.2}^{0.4}$ (0.06)             &                                    \\
        & 0.1-1.9                & ($<0.1$)-1.1              & $<$0.5             &               & 0.2-0.6                     & ($<$0.04)-0.4                      & $<$0.16                                   \\
OCS     & 0.03-0.16              & $\leq$1.6                 & $<$0.22            & 0.1-0.4       & ($<0.02$)-0.06              & $\leq$3.2                          & $<$0.12                            \\
\colrule
\multicolumn{8}{l}{{\bf Possibly identified species$^{\rm h}$:}}\\
HCOOH$^{\rm i}$ & $4_{3}^{5}$ (3) &                             &                    &                & $2.1_{0.9}^{2.5}$ (0.9)       & $2.4_{1.5}^{2.8}$ (0.9)             &                                    \\
        & ($<0.5$)-6             & ($<0.5$)-4                & $<$2               & 0.06-0.14     & ($<0.08$)-2.6               &  ($<0.4$)-7.8                      & $<$4                                   \\

CH$_3$CH$_2$OH$^{\rm i}$& \multicolumn{4}{l|}{$\sim X_{\rm H_2O}$(HCOOH)} & \multicolumn{3}{l}{$\sim X_{\rm H}$(HCOOH)} \\
\multicolumn{8}{c}{}\\
Species & \multicolumn{4}{c|}{$X_{\rm H_2O}^{\rm a}$ [\%]}                                              & \multicolumn{3}{c}{$X_{\rm H}^{\rm b}$ [$10^{-6}$]}\\
        & MYSOs                  & LYSOs                      & BG Stars$^{\rm c}$  & Comets         & MYSOs                       & LYSOs                             & BG Stars$^{\rm c}$ \\        
\colrule
HCOO$^-$$^{\rm j}$& $0.5_{0.5}^{0.7}$ (0.5) &                   &                      &                & $0.18_{0.12}^{0.28}$ (0.14)   &                                   &                                    \\
        & 0.3-1.0                &  $\sim$0.4               & $<$0.1               &                & ($<0.10$)-0.42              & $\sim$0.2                        & $<$0.8                                    \\
CH$_3$CHO$^{\rm j}$& \multicolumn{4}{l|}{$X_{\rm H_2O}$(HCOO$^-$)$\times 11$} & \multicolumn{3}{l}{$X_{\rm H}$(HCOO$^-$)$\times 11$} \\
NH$_4^+$    & $11_{9}^{13}$       & $11_{7}^{15}$              & $8_{6}^{11}$          &                &  $4.1_{1.8}^{5.4}$            & $4.6_{3.2}^{5.8}$                   & $3.8_{2.9}^{4.9}$                    \\
        & 9-34                   & 4-25                      & 4-13               &                 & 1.4-6.0                     & 0.8-12                           & 1.9-9.6                           \\
SO$_2$     & ($<0.9$)-1.4        & $\sim$0.2                 &                    &  0.2            &   $\leq$0.4                 &  $\sim$0.08                       &                                    \\
PAH$^{\rm k}$& $\sim$8            &                           &                    &                 &   $\sim$1                   &                                   &                                    \\
\botrule

\multicolumn{8}{p{17.5cm}}{$^{\rm a}$ Abundances relative to the
  H$_2$O ice column density. For each molecule, the first row gives
  the median and lower and upper quartile values of the detections,
  and in brackets the median including upper limits. The second row
  gives the full range of abundances. If only one row is given, it
  shows the full range. See Table~\ref{t:feat} and \S\ref{sec:abun}
  for references.}\\

\multicolumn{8}{p{17.5cm}}{$^{\rm b}$ As in the $X_{\rm H_2O}$ column,
  but for abundances relative to $N_{\rm H}$.}\\

\multicolumn{8}{p{17.5cm}}{$^{\rm c}$ Quiescent clouds and cores.}\\

\multicolumn{8}{p{17.5cm}}{$^{\rm d}$ Identified based on multiple modes
  or isotopologues in high quality spectra.}\\

\multicolumn{8}{p{17.5cm}}{$^{\rm e}$ The isotopologues $^{13}$CO and
  $^{13}$CO$_2$ were also securely identified, and HDO was possibly
  identified (\S\ref{sec:isot}).}\\

\multicolumn{8}{p{17.5cm}}{$^{\rm f}$ Values derived from the
  simultaneous analysis of multiple features \citep{dar02}.}\\

\multicolumn{8}{p{17.5cm}}{$^{\rm g}$ Identification based on a single
  absorption feature and the profile matches laboratory spectra.}\\

\multicolumn{8}{p{17.5cm}}{$^{\rm h}$ Identification based on a single
  absorption feature and no convincing match to laboratory spectra is
  available.}\\

\multicolumn{8}{p{17.5cm}}{$^{\rm i}$ HCOOH and/or CH$_3$CH$_2$OH may
  be carriers of the 7.24 \mum\ ice band \citep{sch99, obe11}.}\\

\multicolumn{8}{p{17.5cm}}{$^{\rm j}$ HCOO$^-$ and/or CH$_3$CHO may be
  carriers of the 7.41 \mum\ ice band \citep{sch99}.}\\

\multicolumn{8}{p{17.5cm}}{$^{\rm k}$ Using column density of C-C bonds
  derived in \citet{har14} and assuming each PAH species contains 50
  carbon atoms.}\\

\end{longtable}
}
\end{landscape}

Ice abundances relative to elemental hydrogen, $X_{\rm H}$, are
generally calculated by determining $N_{\rm H}=N{\rm (HI)+2{\it
    N}(H_2)}$ from scaling relations with the silicate feature
($\tau_{9.7}$) or the extinction in the near-infrared ($A_{\rm K}$) or
visual ($A_{\rm V}$). These abundances are also given in
Table~\ref{t:abun}, assuming $\tau _{9.7}=0.26 A_{\rm K}$
\citep{boo13} and $N_{\rm H}=1.54\times 10^{22} A_{\rm K}$
\citep{vuo03}. Another way to express ice abundances, of particular
interest to planetary disk models, is the ice-to-rock mass ratio.
Defining ``rock'' as the silicate dust component, the highest observed
ice abundances correspond to ice-to-rock ratios of $\sim$1.5
\citep{pon14}.

Finally, for extragalactic sources only CO, CO$_2$, and H$_2$O ices
were observed in more than a few sightlines. Following the notation of
Table~\ref{t:abun}, $X_{\rm H_2O}$(CO) is $9^{18}_{8}$ (($<4$)-19) for
the LMC \citep{shi10, oli11}, and $<3$ for the SMC
\citep{oli11}. $X_{\rm H_2O}$(CO$_2$) is $17^{25}_{9}$ (6-41) for the
SMC \citep{oli13} and $36^{39}_{29}$ (10-46) for the LMC \citep{shi10,
  sea11, oli09, shi08}. Abundances relative to hydrogen were not
published. The Magellanic Cloud abundances are compared with Galactic
environments in \S\ref{sec:abunvar}.

\subsection{Identification Issues}~\label{sec:iden}

The origin of a number of ice absorption features is uncertain
(Table~\ref{t:feat}).  As opposed to gas phase features \citep{her09},
the identification of ice features is complicated by the dependence of
the peak position on the ice environment (\S\ref{sec:lab} and
\ref{sec:prof}). Therefore, Table~\ref{t:abun} is divided into
categories of ``securely'', ``likely'', and ``possibly'' identified
carriers.  For securely identified species, more than one vibrational
mode was detected and the depth and profile were successfully fitted
with models and laboratory spectra. Although CO ice has only one mode
(4.67 \mum), the $^{13}$CO isotopologue at 4.78 \mum\ was detected as
well. Species in the ``likely identified'' category have only one
securely detected absorption band, but an excellent fit to the profile
and position was demonstrated and the abundance is reasonably well
understood. Species in the ``possibly identified'' category lack a
convincing profile fit or would have an unreasonably high abundances
(e.g., compared to gas phase measurements). SO$_2$ is in this category
as well, because, although the 7.63 \mum\ feature was well fitted by
SO$_2$ in CH$_3$OH ices, it is so far detected in only one line of
sight. In the remainder of this section some identification issues are
highlighted.

\paragraph{The 3.25 \mum\ (PAH?) feature.} 
\marginpar{\textcolor{blue}{\bf PAHs:} Polycyclic Aromatic Hydrocarbon
  molecules, responsible for the ubiquitous infrared emission
  features, and, in dense environments, likely also frozen out in ice
  mantles.}  The 3.25 \mum\ absorption feature depth, measured toward
a handful of YSOs, was found to correlate better with the 9.7 than the
3.0 \mum\ band depth, consistent with a more refractory carrier
\citep{bro99}. The C-H stretch mode of PAH species, detected by its
3.29 \mum\ emission throughout the Galaxy, is thus a good candidate
\citep{sel94}.  Indeed, the strengths of other absorption features in
YSO spectra, in particular in the 5-7 \mum\ region, are consistent
with PAHs embedded in the ices \citep{har14}. No convincing overall
laboratory fit is available, however, due to the heterogeneity of PAH
species, including size distributions and ionization within the ice
\citep{bou11}. An alternative proposed carrier is NH$_4^+$, which is
also less volatile than H$_2$O ices \citep{sch03}, and whose spectrum
also matches the 6.85 \mum\ feature (see below).

\paragraph{The 3.47 \mum\ (NH$_3$) feature.} 
This feature, not to be confused with the 3.40 \mum\ feature due to
saturated hydrocarbons only found in the diffuse ISM, was originally
tentatively attributed to nano-diamonds \citep{all92} in dense clouds.
A very good correlation of the 3.47 and 3.0 \mum\ band depths
indicates a volatile nature of the carrier, however \citep{chi96,
  bro99}. Independent evidence came from spectropolarimetry, showing
that this feature is polarized at a similar strength as the 3.0
\mum\ band \citep{hou96}.  It is consistent with an origin in ammonia
hydrates (\S\ref{sec:30}; \citealt{dar01, dar02}).  This is considered
the preferred carrier, because it also explains much of the overall
long-wavelength wing of the 3.0 \mum\ band and other NH$_3$ features
throughout the infrared spectra of YSOs.

\paragraph{The 4.62 \mum\ (``XCN'') feature. }
The 4.62 \mum\ feature was attributed to a carrier with a C$\equiv$N
(triple bond) group since its first detection
\citep{lac84}. Laboratory experiments in which CO-bearing ices
containing NH$_3$ were photolyzed reproduced this feature
\citep{dhe86} and pointed to the C$\equiv$N bond of a small
molecule. Hence, the observed absorption became known as ``the XCN
feature'', but many laboratory studies have pointed to the true
carrier as the ionic species OCN$^{-}$ (\citealt{dem98} and references
therein). A good correlation between the gas phase HNCO, presumably
photodesorbed at cloud edges, and solid state OCN$^-$ abundances
supports this \citep{obe09a}. Later it was shown that the interstellar
absorption is a composite of features peaking at 4.60 and 4.64
\mum\ \citep{pon03}. Only the peak position and profile of the 4.64
\mum\ component are in agreement with the OCN$^-$ species, most likely
in polar ices that have possibly been heated \citep{bro05}.  It thus
appears that its strength relative to apolar ices is an indicator of
thermal processing, but only for environments that have experienced
the ``catastrophic'' CO freeze out stage in which HNCO and
subsequently OCN$^-$ were formed (\S\ref{sec:modelacc};
\citealt{obe11}).  The nature of the 4.60 \mum\ component is less
clear.  \citet{pon03} suggest CO in a different binding site, and
\citet{obe11} suggest OCN$^-$ in apolar ices although no good fit was
obtained for any mixture considered by \citet{bro05}.

\paragraph{The 6.0 and 6.85 \mum\ (salts, organic residue, PAH, NH$_4^+$?) features. }
Securely identified species do not fully explain the observed
prominent 6.0 and 6.85 \mum\ absorption features in most sightlines
(Fig.~\ref{f:58um}). Dilution of H$_2$O in other species, such as the
abundant CO$_2$, enhances the strength of the H$_2$O bending mode
relative to the 3.0 \mum\ band \citep{kne05}, but this is generally
insufficient to explain all 6.0 \mum\ absorption.  HCOOH could be an
important contributor but its identification in the ices is uncertain
(see below). The carrier of the 6.85 \mum\ feature is not related to
the bending modes of aliphatic hydrocarbons seen at similar
wavelengths in the diffuse ISM due to the absence of the corresponding
stretching modes at 3.40 \mum.  The CO$_3^-$ ion in minerals
(carbonates, e.g., MgCO$_3$ or CaCO$_3$) is also excluded due to the
absence of features at longer wavelengths in the spectra of YSOs
\citep{sch96} and the excellent correlation between the depths of the
6.85 \mum\ absorption and known ice features \citep{tie87a}. The
latter argument also excludes other refractory carriers. The NH$_4^+$
ion in salts \citep{kna82}, is considered the most likely carrier, but
it lacks a convincing profile fit. While the observed strengths
correlates very well with H$_2$O ice \citep{pon06, boo08}, the
laboratory profile is too broad in H$_2$O ices \citep{gal10}. After
H$_2$O ice sublimation, a better fit is obtained, and at even higher
temperature the peak position shifts to longer wavelengths, consistent
with observations of YSOs (enhanced C4/C3 component ratios;
\S\ref{sec:6068}).  Finally, a very broad absorption observed toward a
number of YSOs (C5 component; \S\ref{sec:6068}) has been ascribed to
salts \citep{sch03} and organic residue \citep{gib02}, both of which
are energetic and/or thermal processing products, but these
identifications need independent spectroscopic confirmation.

\paragraph{The 7.24 \mum\ (HCOOH?) feature. }
This feature was attributed to the C-H deformation mode in HCOOH, and
specific mixtures containing both H$_2$O and CH$_3$OH are needed to
match the peak position and profile \citep{sch99}. The depth is
consistent with the much stronger C=O stretch mode at 5.85
\mum\ \citep{sch96}, but its identification was put into question
because of issues with the baseline in the laboratory spectra
\citep{boo08, obe11} and an abundance that is a factor of $10^4$
larger compared to the gas phase in hot cores is also not
understood \citep{bis07}. \citet{obe11} propose pure ethanol instead.
Other species tend to have slightly blue-shifted peak positions
(HCONH$_2$, C$_5$H$_{12}$; \citealt{sch99}; NH$_2$CH$_2$OH;
\citealt{bos09}) and are listed in Table~\ref{t:abunuplim}.

\subsection{Upper Limits}~\label{sec:uplim}

Table~\ref{t:abunuplim} shows abundance upper limits of
astrophysically relevant species. These were determined by comparing
laboratory spectra with the spectra of YSOs and background stars,
taking into account position and profile dependencies on the ice
composition. For symmetric species (in particular O$_2$, N$_2$, and
H$_2$), interactions in the ice matrix induce a dipole moment, and the
band strength is a very strong function of molecular
environment. These make the derived abundance upper limits
uncertain. Indirect measurements of the O$_2$ and N$_2$ abundances in
mixtures with CO by the effect on the $^{12}$CO \citep{els97} and
$^{13}$CO \citep{boo02a} band profiles lower the upper limits for
N$_2$ by a factor of $\sim 5$, but do not further constrain those of
O$_2$.  The case of H$_2$ refers to energetically processed ices that
are able to retain the produced H$_2$ \citep{san93}. In pure form
H$_2$ sublimates at 3 K, well below the temperature of dense
clouds. An upper limit rather than the detection claimed in
\citet{san93} is quoted because of confusion with photospheric
absorption lines at 2.415 \mum.

\newpage

\begin{longtable}{llll|l}
\caption{Abundance upper limits of astrochemically relevant ices and
  corresponding detections in comets}\label{t:abunuplim}\\
\toprule
Species                   & $X_{\rm H_2O}$       & $X_{\rm H}$        & Environment$^{\rm a}$ (reference)$^{\rm b}$ & $X_{\rm H_2O}^{\rm c}$ (comet)\\
                          & \%                 & $10^{-6}$         &                                          & \%                 \\
\colrule
N$_2$                     & $<0.2-60$         & $<0.1-28$        & Taurus cloud (1, 2, 3)                         &                    \\
O$_2$                     & $<39$              & $<60$            & LYSO R CrA IRS2 (4)                      &                    \\
                          & $<15$              & $<30$            & MYSO NGC 7538 IRS9 (4)                    &                    \\
H$_2$                     & $<68$              & $<14$             & LYSO WL 5 (5)                             &                    \\
H$_2$S                    & $<0.3-1$           & $<0.04-0.12$     & MYSOs (6)                                & 0.12-1.4           \\
                          & $<1-3$             & $<0.6-1.6$       & Taurus cloud (6)                         &                    \\
H$_2$O$_2$                & $<2-17$            & $<0.6-8$         & YSOs, Taurus cloud (7)                   &                    \\
C$_2$H$_2$                & $<1-10$            & $<0.4-4$         & MYSOs (8)                                & 0.1-0.5            \\
C$_2$H$_6$                & $<0.3$             & $<0.14$          & MYSO NGC 7539 IRS9 (8)                   & 0.1-2              \\
C$_5$H$_{12}$              & $<15$              & $<10$            & MYSO W 33A (12)                           &                  \\
C$_3$O$_2$                & $<5$               & $<2$             & YSOs (9)                                 &                    \\
N$_2$H$_4$, N$_2$H$_5^+$   & $<10$              & $<4$             & MYSOs (8)                                &                    \\
HNCO                      & $<0.3-0.7$         & $<0.10-0.24$     & MYSOs (11)                                & 0.02-0.1           \\
HCONH$_2$                 & $<1.5$             & $<1$             & MYSO W 33A (12)                           & 0.002                 \\
NH$_2$CH$_2$OH            & $<3-6$             & $<2-4$           & MYSO W 33A (13)                           &                 \\
NH$_2$CH$_2$COOH$^{\rm d}$  & $<0.3$             & $<0.1$           & MYSO W 33A (10)                            &                    \\
\botrule
\multicolumn{5}{p{16cm}}{$^{\rm a}$ The environment for which the
  upper limit was determined.}\\
\multicolumn{5}{p{16cm}}{$^{\rm b}$ References:
(1)-\citealt{els97},
(2)-\citealt{san01},
(3)-\citealt{boo02a},
(4)-\citealt{van99},
(5)-\citealt{san93},
(6)-\citealt{smi91},
(7)-\citealt{smi11},
(8)-\citealt{bou98},
(9)-\citealt{pon08},
(10)-\citealt{gib04},
(11)-\citealt{bro05},
(12)-\citealt{sch99},
(13)-\citealt{bos09}
}\\
\multicolumn{5}{p{16cm}}{$^{\rm c}$ Detections toward comets
  \citep{mum11}.}\\
\multicolumn{5}{p{16cm}}{$^{\rm d}$ Glycine}\\
\end{longtable}

Most species in Table~\ref{t:abunuplim} were studied because of their
expected production in the gas phase or on grain surfaces. C$_3$O$_2$
is a by-product of the energetic formation of CO$_2$
(\S\ref{sec:modelen}).  Species for which, to the best of our
knowledge, no detections were claimed or upper limits were determined
include the electronically stable ``generation 0'' species whose
formation was studied in the laboratory and are listed by
\citet{the13}: HCN, CH$_3$NH$_2$, CH$_2$NH, NH$_2$OH, NH$_2$CHO,
CH$_3$CHO, and HNCO. Other relevant species include O$_3$, HOCN, and
H$_2$CO$_3$. Energetic processing and thermal reactions efficiently
produce more complex species, but these are increasingly hard to
detect because of a lack of distinct absorption features, as evidenced
by the high upper limit of glycine listed in
Table~\ref{t:abunuplim}. Millimeter wave spectroscopy of gas phase
species is orders of magnitude more sensitive, and must be used to
determine their abundances after sublimation off the grains.

\subsection{Abundance Variations}~\label{sec:abunvar}

A recurring conclusion in the literature is that the ice abundances
relative to H$_2$O are to the first order similar over a wide range of
environments. As sample sizes and uncertainties have improved, clear
trends of variations by factors of 2-10 were discerned
(Table~\ref{t:abun}).  Most of these relate to CO depletion and
subsequent chemical reactions (\S\ref{sec:modelacc}).  A statistical
study by \citet{obe11} revealed that $X_{\rm H_2O}$(CO) and $X_{\rm
  H_2O}$(CO$_2$) are factors of $\sim$3 and 1.5 larger toward LYSOs
and the quiescent regions of isolated cores compared to MYSOs and
Taurus background stars. Within samples of MYSOs and LYSOs, $X_{\rm
  H_2O}$(CH$_3$OH) and $X_{\rm H_2O}$(OCN$^-$) (4.64 \mum\ component)
vary by an order of magnitude. CH$_3$CH$_2$OH or HCOOH fall in this
category too. In quiescent regions, the observed $X_{\rm
  H_2O}$(CH$_3$OH) varies from $<$3\% (Taurus) to $\sim$12\% in some
isolated dense cores.

\begin{figure}[t!]
\centerline{\psfig{figure=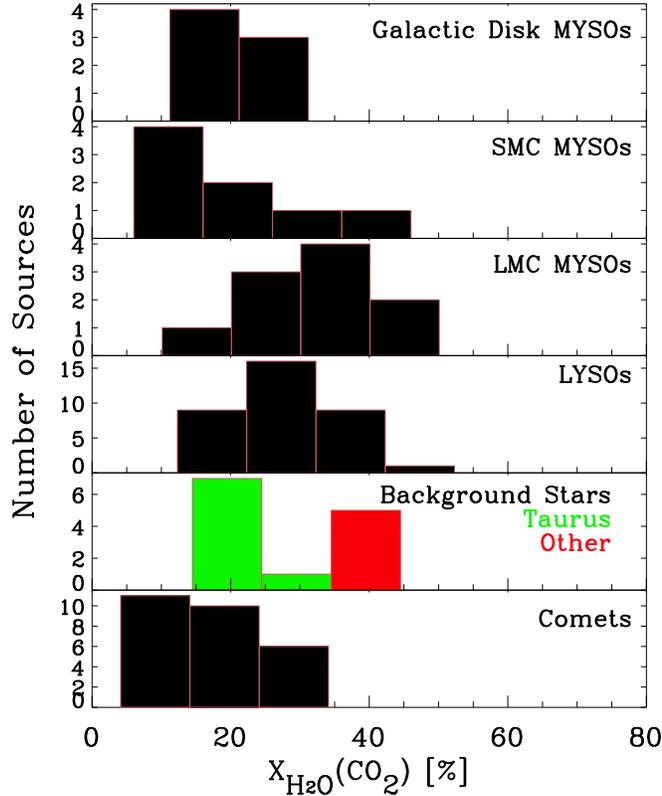,height=25pc,angle=90}}
\caption{CO$_2$ abundances relative to H$_2$O in a range of
  environments. For the comets, the abundance represents the ratio of
  gas phase production rates (within the H$_2$O sublimation radius at
  2.7 AU; \citealt{oot12}). For background stars, Taurus has
  distinctly lower abundances than other regions.}\label{f:co2h2o}
\end{figure}

The overview by \citet{obe11} does not include extragalactic and GC
ices, which are particularly well studied in
CO$_2$. Figure~\ref{f:co2h2o} shows a comparison of $X_{\rm
  H_2O}$(CO$_2$) in all environments with accurate measurements. It
highlights the systematic differences between environments, which are
often rather small, but significant.  MYSOs in the Galactic Disk have
abundances that are typically a factor of 1.5-2 smaller compared to
those in the LMC and compared to Galactic LYSOs. SMC MYSOs have low
abundances when the H$_2$O column is low, but converge to the LMC
values at higher columns \citep{oli13}. For background stars, the
distribution is double-peaked. Low abundances ($\sim$20\%) are
observed toward seven stars behind Taurus \citep{whi09}, while high
abundances ($\sim$40\%) are found for five lines of sight behind
the Serpens \citep{kne05} and Lupus \citep{boo13} clouds, and the IC
5146 \citep{whi09}, L429-C, and L483 cores \citep{boo11}. The
discriminator appears to be the line of sight extinction: all but one
of the Taurus sightlines have \av$<25$ mag, while all but one of the
other sightlines have \av$>25$ mag. It may thus be related to CO
freeze out (\S\ref{sec:modelacc}).

\begin{landscape}
\begin{figure}[t!]
\centerline{
\psfig{figure=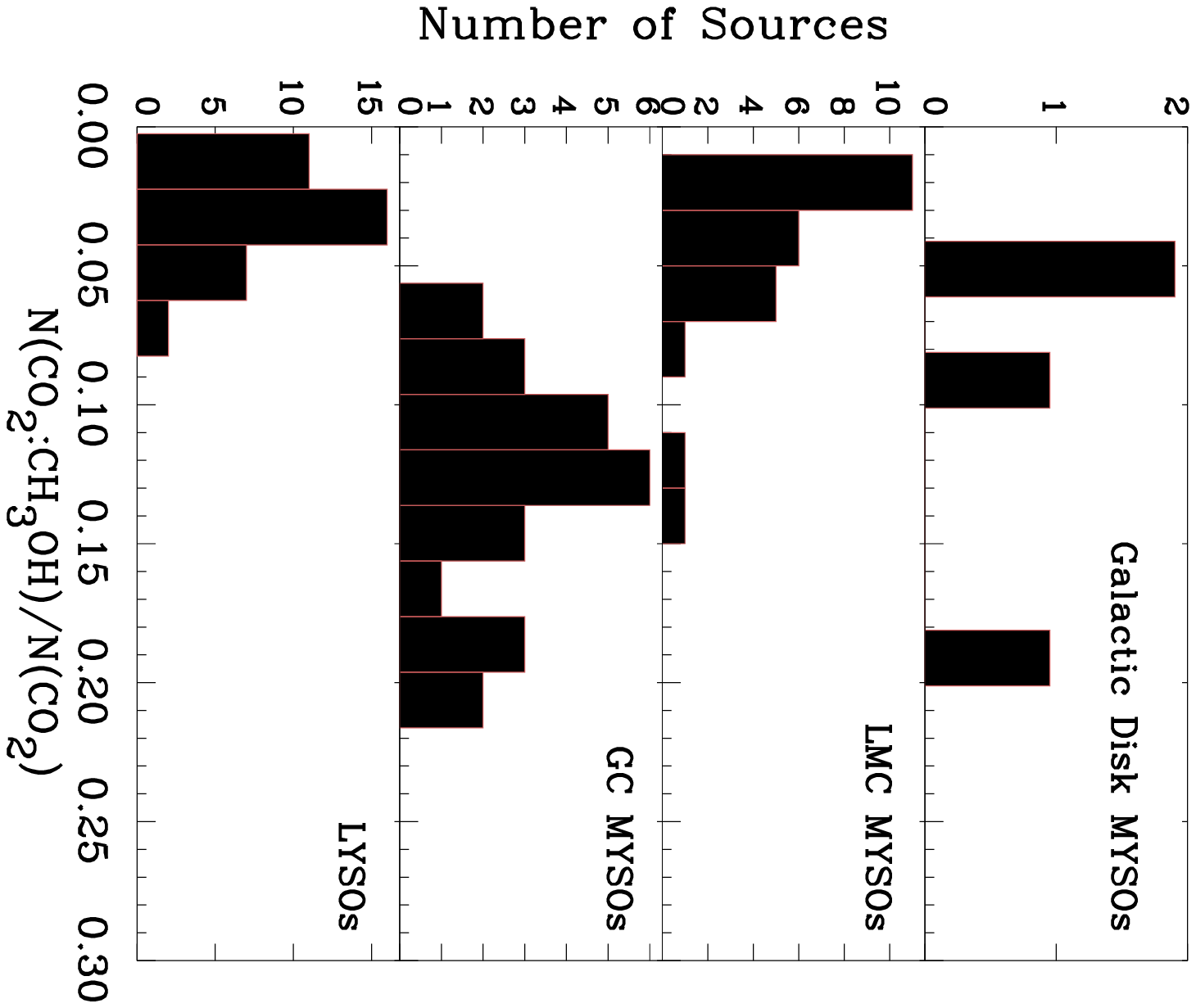,height=21pc,angle=90}
\psfig{figure=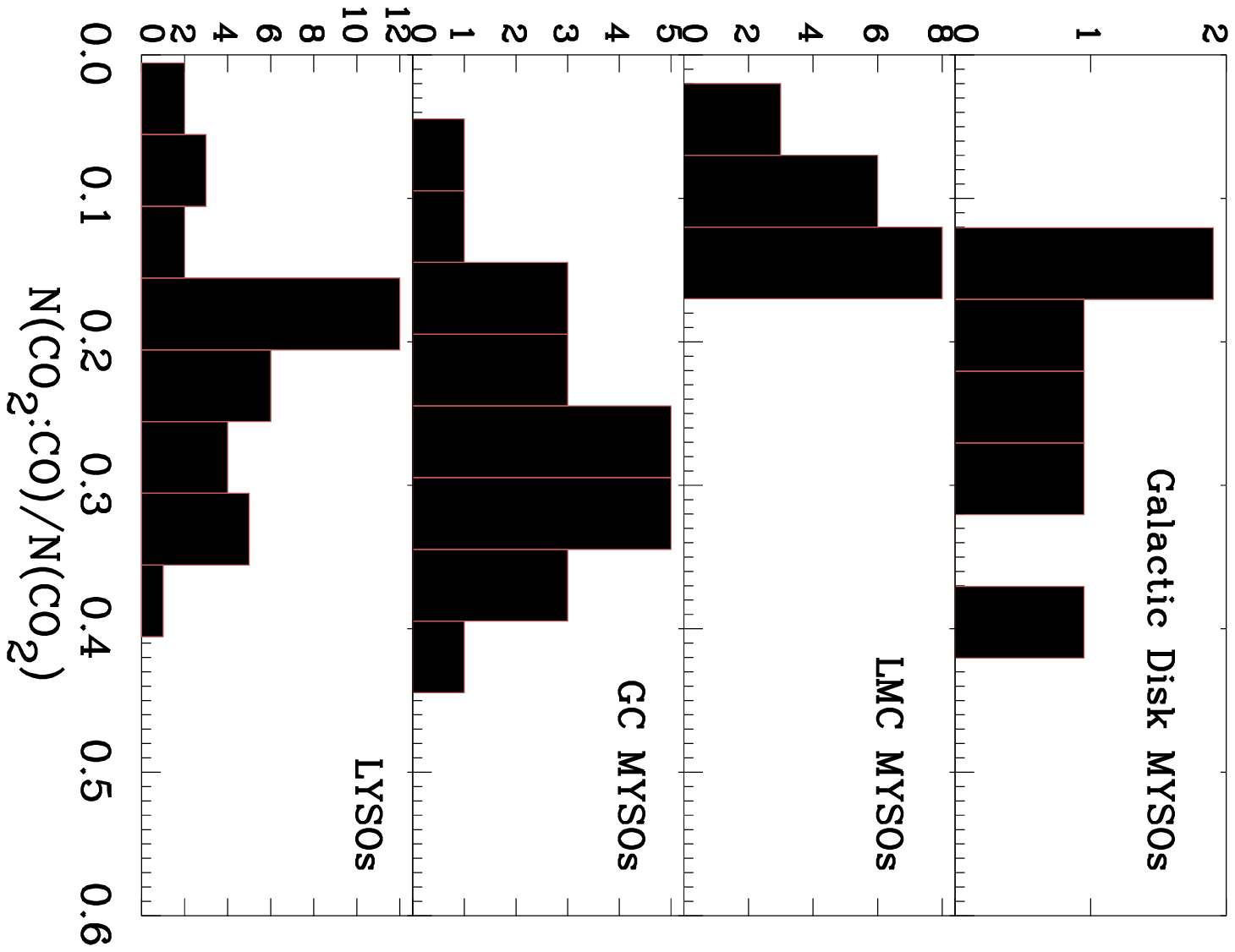,height=21pc,angle=90}
\psfig{figure=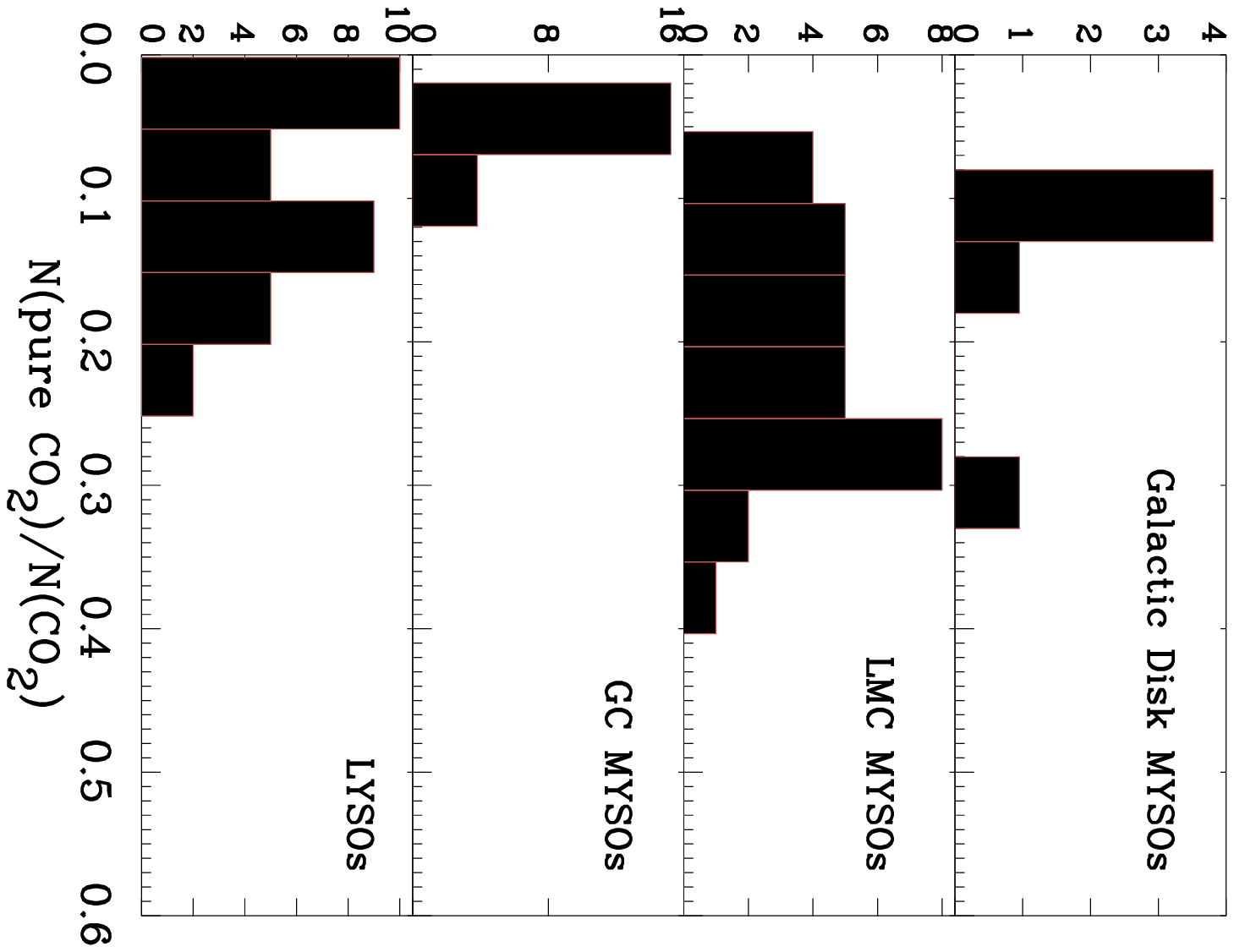,height=21pc,angle=90}
}
\caption{The 15.2 \mum\ band of CO$_2$ shows distinct profile
  variations.  Decompositions following \citet{pon08} show that the
  CO$_2$:CH$_3$OH and CO$_2$:CO components are particularly strong
  toward the GC MYSOs. LYSOs show a particular lack of
  CO$_2$:CH$_3$OH, while MYSOs in the LMC show an enhancement in pure
  CO$_2$. Not shown is the CO$_2$:H$_2$O component, for which median
  values are quoted in \S\ref{sec:abunvar}.}\label{f:co2decomp}
\end{figure}
\end{landscape}

The 15.2 \mum\ CO$_2$ band profiles vary between environments as
well. The 5-component fits applied to LYSOs and Galactic MYSOs
\citep{pon08} were also applied to MYSOs in the GC CMZ \citep{an11}
and in the LMC \citep{sea11}.  The GC MYSOs have twice as much
CH$_3$OH:CO$_2$ and CO:CO$_2$ ices relative to other environments
(Fig.~\ref{f:co2decomp}). LMC MYSOs, on the other hand, have much
stronger pure CO$_2$ components, indicating a larger degree of
processing.  Not shown here is the dominant CO$_2$:H$_2$O component,
which is somewhat less abundant for the GC MYSOs and LMC MYSOs ($\sim
55$\%) compared to Galactic LYSOs ($\sim 70$\%).  For the LMC sources,
the sum of the pure CO$_2$ and CO$_2$:H$_2$O ices is similar to the
polar fraction toward LYSOs \citep{sea11}, suggesting that the pure
CO$_2$ ices are ``tapped'' from the polar ices upon heating. For GC
MYSOs, the CH$_3$OH:CO$_2$ component makes up for the difference with
LYSOs. Finally, some LYSOs have lower (CO$_2$:H$_2$O)/CO$_2$ fractions
and higher $X_{\rm H_2O}$(CO$_2$), which is due to enhanced CO$_2$
formation in CO-rich environments (e.g., source SVS 4-10 in
Fig.~\ref{f:15um}).

Variations of ice abundances relative to $N_{\rm H}$ are evident too.
The CO depletion in nearby quiescent clouds is incomplete (40-50\%;
\citealt{whi85, chi94}) and the H$_2$O abundance is also a factor of
$\sim$2 below the values seen toward embedded YSOs \citep{boo13}. At
higher densities ($10^5-10^6$ \cubcm) in cores and envelopes all ice
abundances increase, but especially those of CO and chemically related
species (CH$_3$OH, CO$_2$, OCN$^-$; \citealt{pon04}).

\subsection{Isotopologues}~\label{sec:isot}

The $^{13}$CO and $^{13}$CO$_2$ isotopologues were securely detected
toward interstellar sources, and HDO tentatively.  $^{12}$CO/$^{13}$CO
and $^{12}$CO$_2$/$^{13}$CO$_2$ ratios of 68-71 (two sightlines;
\citealt{boo02a, pon03}) and 52-110 (thirteen sightlines;
\citealt{boo00a}) were derived, respectively. The
$^{12}$CO$_2$/$^{13}$CO$_2$ ratios increase with Galactocentric
radius, in accordance with the elemental isotopic ratios.  HDO was
tentatively detected at an abundance of 2-22\% relative to H$_2$O
toward LYSOs with edge-on disks \citep{aik12}. Otherwise, low upper
limits were derived: $<$0.2-1\% for intermediate mass and massive YSOs
\citep{dar03}, and $<$0.5-2\% for LYSOs \citep{par03}. The relatively
low deuterium fractionation of H$_2$O is consistent with formation in
an early cloud phase (\citealt{cec14}; \S\ref{sec:modelacc}).

\subsection{Interstellar Versus Cometary Ice Abundances}~\label{sec:abunss}

Nearly all securely, likely, and possibly detected ice species outside
the solar system are also detected toward comets
(Table~\ref{t:abun}). Cometary ices are indirectly traced after
outgassing and ions have recombined to their neutral
counterparts. Interstellar NH$_4^+$ should thus be compared to
cometary NH$_3$, OCN$^-$ to HNCO, and HCOO$^-$ to HCOOH. Overall, the
median abundances of all C- and N-bearing species relative to H$_2$O
in comets are below those of LYSOs.  For CH$_4$, CH$_3$OH, and CO, the
distributions overlap, however \citep{obe11}.  Using the measurements
in comets from \citet{oot12}, a similar conclusion can be drawn for
CO$_2$ ices (Fig.~\ref{f:co2h2o}): the median abundance is well below
that of LYSOs (the upper quartile of comets is similar to the lower
quartile of LYSOs), but there is some overlap. For CH$_3$OH, it
appears that the overlap is caused by the class of organic-enriched
comets, which has members from both the Oort Cloud and the Kuiper
Belt, and has abundances similar to the median of LYSOs
\citep{mum11}. There is no overlap in the distribution of NH$_3$
abundances, however. The discrepancy is even larger when the NH$_3$
and NH$_4^+$ abundances toward LYSOs are added.

Because cometary ices are traced indirectly using gas phase rotational
transitions, more sensitive abundance measurements can be made
compared to ice measurements toward LYSOs. This is evident in
Table~\ref{t:abunuplim}, which shows that upper limits for H$_2$S,
C$_2$H$_2$, C$_2$H$_6$, and HNCO toward LYSOs are all above the
detections for comets.  Considering that HNCO in comets may directly
trace OCN$^-$ in LYSOs, the trend of low cometary C and N abundances
are confirmed.  The median is well below that of LYSOs, but there is
some overlap.  Finally, a number of species were detected toward
comets, and no measurements are currently available for LYSO ices: HCN
(0.08-0.5\%), S$_2$ (0.001-0.25\%), HOCH$_2$CH$_2$OH (0.25\%),
HCOOCH$_3$ (0.09\%), HNC (0.003-0.05\%), CH$_3$CN (0.008-0.04\%),
HC$_3$N (0.003-0.07\%), and H$_2$CS (0.05\%). In view of the uncertain
identification of a number of interstellar ice features, in particular
those in the 5-8 \mum\ wavelength region, a search for these ices is
warranted.

\subsection{Elemental Budget}~\label{sec:elem}

The budget of chemical elements constrains the identification of
detected features and helps direct searches for new volatile and
refractory species.  The securely-, likely-, and possibly-detected
ices do not nearly account for the abundance of the heavy elements O,
C, N, and S that is presumably available to be included in volatiles
(i.e, excluding depletion in known refractory dust and gas phase
volatiles). For O, $\sim 35$\% is ``missing'', for C $\sim$30\%, for N
70-85\%, and for S $\sim$95\%.

The total amount of the available O included in H$_2$O, CO, and CO$_2$
ices is similar in different quiescent clouds ($\sim$26\%), even
though the abundances of the individual species vary \citep{whi09}.
\citet{obe11} find a similar median value for MYSOs, and a somewhat
larger median value of 34\% for LYSOs, with a maximum of 61\% in
regions of high CO freeze out.  There is lingering uncertainty
concerning putative additional carriers \citep{whi10}. Solid O$_2$
could account for up to 10\% of O based on the limit in \citet{van99},
and the possibly identified species HCOOH for another few percent.

Of the C budget presumably available for inclusion in volatiles, 27
and 14\% is taken up by known ices in LYSOs and MYSOs, respectively
\citep{obe11}. Maximum values of a factor of 2 larger are
measured. Frozen PAH species are likely an important reservoir of
(refractory) carbon in the ices, at 18\% of the volatile budget, at
least, using absolute band strength measurements \citep{har14} of
neutral pyrene, but this needs to be re-assessed when the degree of
ionization of the frozen PAHs is constrained \citep{kea01b, bou11}.

NH$_3$, NH$_4^+$, and OCN$^-$ account for $\sim$11\% of the available
N \citep{obe11}. It would be half of that if the identifications of
NH$_4^+$ and OCN$^-$ are incorrect. In some sightlines, up to 30\%
could be included in the ices.  It is plausible that much of the
``missing'' N is in the form of solid N$_2$, but this is extremely
difficult to confirm (\S\ref{sec:uplim}).  The (uncertain) upper
limits to the solid N$_2$ abundance (Table~\ref{t:abunuplim})
correspond to about 40\% of the cosmic N budget.

Finally, OCS and, if its identification is confirmed, SO$_2$,
contribute only 4\% to the available sulfur budget \citep{pal97}.  The
upper limit to H$_2$S (Table~\ref{t:abunuplim}) would account for
another percent.

\section{GRAIN SIZE AND MANTLE THICKNESS}~\label{sec:size}

Mantle growth is expected to proceed at a rate independent of the
grain size, although the smallest grains ($<0.002$ \mum) will be
ice-less as they become too hot upon stochastic photon heating
\citep{hol09}. The elemental budget (\S\ref{sec:elem}) gives upper
limits to the mantle thickness. For H$_2$O-rich ices this is 0.005
\mum\ assuming an MRN grain radius $a$ distribution, at which point
all O not included in gaseous CO and silicates is presumed to be
incorporated into H$_2$O. Subsequent complete freeze-out of CO
increases the mantle thickness by a further factor $\sim$2. Such
mantle growth can thus not significantly increase the sizes of the
largest grains ($a>$0.1 \mum), but it can lead to a large fractional
increase in the sizes of the abundant small grains that represent most
of the surface area of the dust.  Coagulation of the sticky icy grains
will lead to further growth in dense molecular clouds \citep{orm09}.

The observed ice features place some constraints on the grain size and
mantle thickness.  There is strong evidence for grains that are small
compared to the wavelength ($a<<\lambda/2\pi$) with comparatively
thick ice mantles.  The peak optical depth of the H$_2$O stretch mode
is generally centered at 3.0 \mum\ ($a<<0.5$ \mum), and the profiles
of the CO and CO$_2$ stretch modes are single peaked.  Thin mantled
grains (core over mantle volume ratios larger than $\sim 0.2$;
\citealt{tie91}) would have double-peaked profiles.  In addition,
spectrally resolved polarimetry of the 3.0 \mum\ band shows a shift of
the polarization peak to longer wavelengths indicative of dichroic
polarization (alignment of elongated grains by magnetic fields;
\citealt{whi11a}; Andersson et al., this volume) by grains smaller
than $\sim$0.1 \mum; \citep{hag83, hou96}.  To test the presence of
ices on grains much smaller than this, observations of the electronic
transition of H$_2$O at 0.14 \mum\ would be needed.

There is evidence for icy grains that are large compared to the
wavelength as well, but with the exception of some clear-cut cases
(e.g., abundant $\sim 0.8$ \mum\ icy grains in an LYSO disk;
\S\ref{sec:disks}; \citealt{ter12a}) it has encountered problems. Mie
scattering calculations show that grain sizes of up to $\sim$0.5
\mum\ would be needed to explain the long wavelength wing of the 3.0
\mum\ ice band (\S\ref{sec:30}). Such large grains are, however,
inconsistent with the spectropolarimetric measurements mentioned
above. Also, in models satisfying all observational constraints (grain
size distribution, elemental budget, ice band profiles) \citet{smi93}
find that their observations of the 3.0 \mum\ profile in Taurus can
only be explained if ice mantles are preferentially present on
specific grain populations (e.g., on larger grains or on silicate
grains).

Indirect evidence for a distinct population of much larger,
micron-sized icy grains comes from diffuse scattered light observed
toward dense cloud cores (``coreshine''; \citealt{and14}).  Such
grains were also suggested as an alternative explanation of the
long-wavelength wing of the 4.67 \mum\ CO ice band (\S\ref{sec:467};
\citealt{dar06}). Large grains have been suggested as a source of the
``missing'' oxygen (\S\ref{sec:elem}; \citealt{jen09}), because for
radii larger than a few micron, the infrared absorption profiles
become too wide and shallow to be detected.  In light of the small
variations of the 3.0 \mum\ band profile (\S\ref{sec:30}), these
suggestions need further study, including an assessment of the effects
of grain coagulation on the band profiles.

\section{CONSTRAINTS ON ICE EVOLUTION MODELS}\label{sec:model}

\begin{landscape}
\begin{figure}
\centerline{\psfig{figure=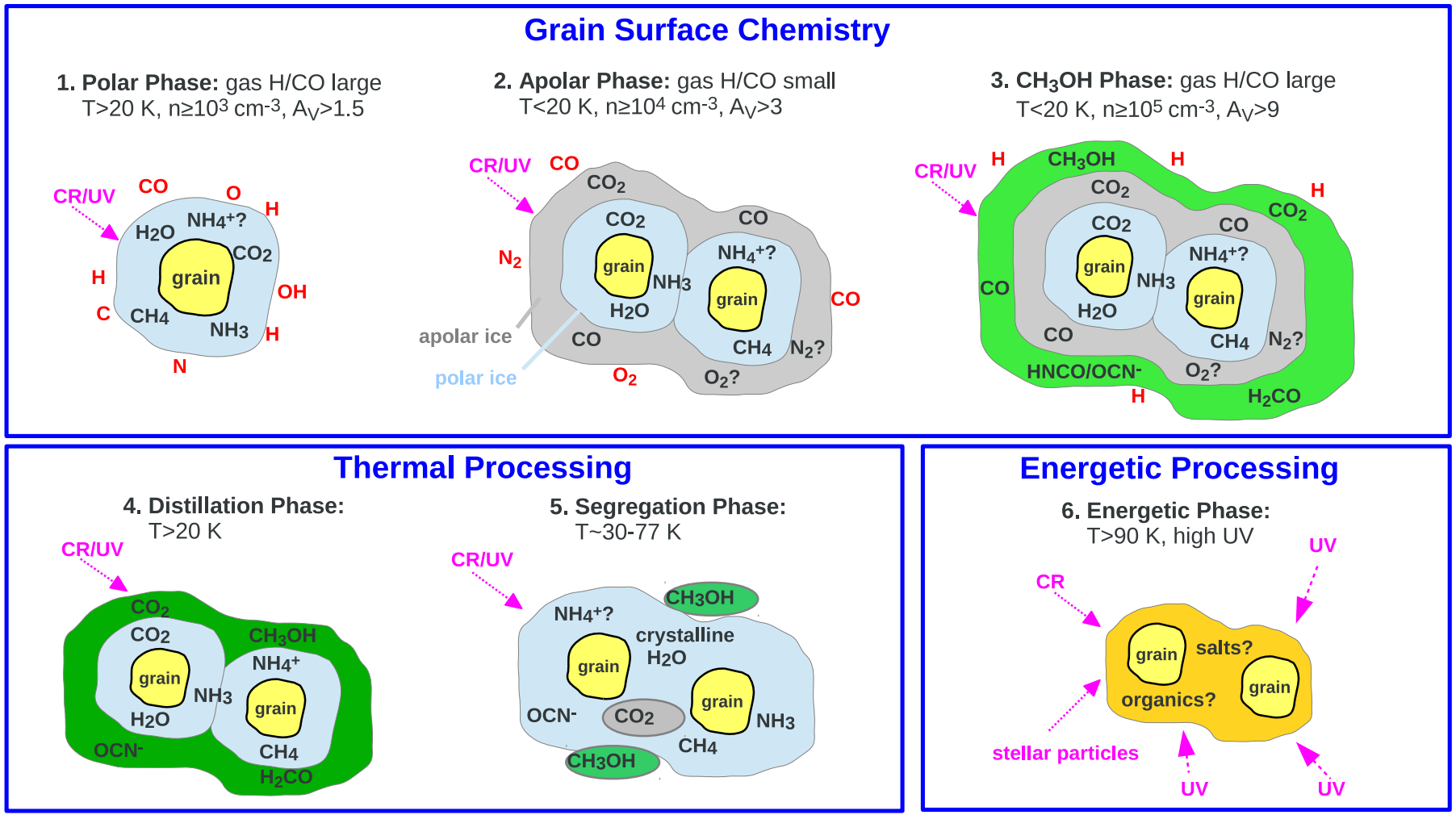,width=60pc}}
\caption{Schematic view of the observationally constrained composition
  of icy grain mantles, highlighting the physical and chemical
  processes at play. Six phases are shown, which could represent
  progressive evolutionary phases in quiescent clouds (phases 1-3) and
  YSO envelopes and disks (phases 3-6). Gas phase species
  participating in grain surface reactions are indicated in red.  In
  phases 1-5, the ices experience CR and secondary UV fields that only
  minimally affect the main ice species. As the ice temperature rises
  (phases 4 and 5), thermal reactions or reactions involving radicals
  may proceed. Stage 6 is least observationally constrained. The
  relative grain and mantle sizes in the first 5 stages are realistic
  (maximum mantle thickness of 50 \AA\ and minimal grain core sizes of
  20 \AA). After the initial ice layers are formed (phase 1), grain
  coagulation accelerates.}\label{f:mantle}
\end{figure}
\end{landscape}

The commonalities and differences in the ice abundances and band
profiles observed over a wide range of environments provide strong
constraints on models of ice formation and evolution.  We do not
review the models here, but rather summarize the observational
constraints. Figure~\ref{f:mantle} illustrates the probable scenario.

\subsection{Accretion and Grain Surface Chemistry}~\label{sec:modelacc}

The observed ice abundances are generally in good agreement with
models of accretion from the gas phase followed by grain surface
chemistry.  In particular, the efficiency of hydrogenation as a result
of the mobility and tunneling capacity of H atoms (e.g.,
\citealt{tie82}) is confirmed by the observations.  The dominance of
H$_2$O ice and mixtures with H$_2$O in all Galactic and extragalactic
environments points to a phase of molecule formation on cold grain
surfaces at which the accreting gas is atomic H- and O-rich.  This is
a phase of relatively low densities ($n\geq 10^3$ \cubcm) early in the
cloud evolution (H/H$_2$ ratios decrease rapidly with density;
\citealt{hol71}).  The ice formation threshold of \av=1.6 mag near
local cloud edges is understood as the onset of rapid ice mantle
growth after a monolayer has formed and the effect of photodesorption
decreases with increasing extinction by dust into the cloud
(\av$\propto ln(G_{\rm 0}/n)$; \citealt{hol09}).  This phase also
includes the formation of CH$_4$, CO$_2$ (polar component, likely via
CO+OH; \citealt{iop11}), and NH$_3$ which are all intimately mixed
with H$_2$O.  Variations in CO$_2$/H$_2$O ratios in this early phase
reflect variations of the CO accretion rates. On one hand, if the gas
phase formation of CO is incomplete, e.g., at low extinctions, more O
can be included in H$_2$O \citep{whi09, obe11}. It may explain the
possibly CO$_2$-free region in the SMC \citep{oli13} and CH$_4$-rich
region in Galactic cloud edges \citep{obe11}.  On the other hand, CO
accretion rates increase at lower temperatures and at higher
densities, leading to enhanced formation of CO$_2$ (up to a factor of
2), still well mixed with the co-formed H$_2$O species. This phase is
CH$_3$OH-poor and has a relatively low CO freeze out ($<50\%$). At the
same time, the presence of the 6.85 \mum\ feature and some absorption
underlying the 6.0 \mum\ band (Fig.~\ref{f:58um}) show that these
early ices are already quite complex. They are formed on a short time
scale ($10^9/n \sim 10^5$ yr at $n=10^4$ cm$^{-3}$; e.g.,
\citealt{hol09}) compared to the cloud life time ($\sim 10^7$ yr) and
compared to the formation time of new species by interstellar CRs or
CR-induced UV fields (\S\ref{sec:modelen}).

Strong enhancements (up to a factor of 10) of the abundances of
CH$_3$OH, OCN$^-$, the carriers of the 7.24 and 7.41 \mum\ features
(CH$_3$CH$_2$OH or HCOOH and CH$_3$CHO or HCOO$^-$), and the apolar
components of the CO and CO$_2$ ices highlight the subsequent phase of
ice formation. It is the episode of ``catastrophic'' CO freeze out at
low temperatures and high densities ($n\geq 10^5$ \cubcm), when the
gas phase H/CO ratios are drastically enhanced, even though the
H/H$_2$ ratios are reduced.  This then leads to H$_2$CO and CH$_3$OH
on a short time scale of $\sim {\rm few}\times 10^4$ years, as
demonstrated in Monte Carlo simulations \citep{cup09}, and likely to a
host of other CO-derived species, including HNCO \citep{fed15} and
CO$_2$ (in the apolar component).  The best observational evidence of
this phase comes from the measurements of ice abundance gradients in
dense cores and Class 0 LYSOs \citep{pon04, pon06}.  The quiescent
regions in the Taurus cloud have not entered this phase yet
(Fig.~\ref{f:co2h2o}), in contrast to a subset of dense cores,
prestellar cores, and LYSOs (e.g., SVS 4-10 in Fig.~\ref{f:15um} has a
CO$_2$/H$_2$O ratio of 50\% and a strong CO:CO$_2$ component;
\citealt{pon03}). Clearly, a complex mixture of ices is formed before
stars form. Apparently, the GC MYSOs (prominent apolar CO$_2$ and
CH$_3$OH:CO$_2$ components) have gone through this phase, but the LMC
MYSOs have not (weak apolar CO$_2$ and CH$_3$OH:CO$_2$ components and
low solid CO abundances). We speculate that high and low metallicities
in these distinct environments, respectively, play a role. High
metallicities reduce the H/CO ratios and increase extinction and thus
CO freeze out.

\subsection{Thermal Processing}~\label{sec:modeltherm}

Observations show that after the formation of the bulk of the ices
through accretion and grain surface chemistry, the evolution of the
ices in the circumstellar envelopes and disks of YSOs is dominated by
thermal processes (\S\ref{sec:prof}).  In summary, sublimation of
apolar ices occurs at temperatures of $\sim 20$ K.  As CO is
``distilled'' out of the CO$_2$ ices, this develops the 15.11 and
15.26 \mum\ absorption peaks of pure CO$_2$. Segregation of CO$_2$
from H$_2$O-rich ices occurs at higher temperatures, starting at 30 K
for the surface layers and thin mantles \citep{obe09b}, and up to
$\sim$77 K for thick mantles \citep{boo00a} at which the H$_2$O
crystallization process starts as well.  This is observed by a
long-wavelength shift and narrowing of the 3.0 \mum\ band, and again
by the development of the 15.11 and 15.26 \mum\ absorption peaks. The
``pure'' CO$_2$ will sublimate at temperatures in the range of
$\sim$45 K for ``distilled ices'', up to the H$_2$O sublimation
temperature of $\sim$90 K for segregated ices, depending on the mantle
thickness \citep{fay11}.  Sublimation of the polar, H$_2$O-rich ices
is traced by increased gas/solid state column density ratios
\citep{boo03}.  The strength of the C4 component of the 6.85
\mum\ feature is apparently enhanced at temperatures above the H$_2$O
sublimation temperature \citep{boo08}.

The sublimation, segregation, and crystallization temperatures quoted
in this review apply to the long time scales in the interstellar
environments.  At any temperature, the probability that the energy
barrier for these processes is surmounted is set by the Boltzmann
distribution.  This translates to time scales of $\nu_0^{-1}{\rm
  e}^{E_{\rm barrier}/T}$ \citep{tie87b}.  For example, for H$_2$O,
with an energy barrier of $E_{\rm barrier}=5500$~K and a frequency
$\nu _0=5\times 10^{-13}~{\rm s^{-1}}$ (O-H bend vibration mode), an
observed sublimation temperature of 180 K at a laboratory time scale
of minutes corresponds to 90 K at space time scales of 10,000 years.
These arguments were used to link the range of pure CO$_2$ peak
strengths in a sample of MYSOs to a time scale of $\sim 3\times 10^4$
yrs needed to heat most of their envelopes to temperatures of $\sim
77$ K \citep{boo00a}. Models show that envelope heating and dispersal
occur at the same time \citep{tak00}.  For low mass envelopes,
dynamical time scales were coupled to temperatures, which is critical
to locate the various processes \citep{pon08}. For a 1 $M_\odot$, 3
$L_\odot$ star, this translates to an apolar ice sublimation radius of
1000 AU, a crystallization and segregation region between 100 and 130
AU, and polar ice sublimation within 100 AU.  In circumstellar disks
surrounding similar stars, the observed ice crystallization
(\S\ref{sec:disks}) is also thought to happen at radii $\geq 100$ AU,
but only in their super-heated optically thin surfaces \citep{chi01}.

For objects of very low luminosity, envelope heating is insufficient
to cause crystallization and segregation \citep{kim12}.  The double
peaked CO$_2$ profile is detected toward $<0.1$ L$_\odot$ YSOs,
however.  It is suggested that the apolar CO$_2$ component is
responsible for that, in combination with heating spikes caused by
episodic accretion events.  During long intervals between the
accretion events, CO$_2$ is formed in the CO-rich ice by reactions
with OH. The heating event then sublimates the CO, and leaves pure
CO$_2$ behind. In a separate study, a CO$_2$ bending mode consisting
of nearly pure CO$_2$ is found toward the 1 $L_\odot$ YSO HOPS 68
(Fig.~\ref{f:15um}; \citealt{pot13}). This may also be interpreted as
a result of episodic heating, but in this case the event is much
stronger (up to 100 $L_\odot$, as in FU Ori objects) and all ices are
sublimated and then recondensed in the order of their condensation
temperature. Fortuitous timing and line of sight geometry inside a
flattened envelope are required to trace this particular ice
component.

The common process of ice heating most likely affects the chemistry as
well, although from an observational point of view there is little
direct evidence for this. A candidate feature produced upon heating is
the ``C5'' component (Fig.~\ref{f:58um}). We speculate that it might
be formed when the radicals, created by energetic processes over the
cloud and envelope life time, become more mobile and react at higher
temperatures \citep{gar08, vas14}.  Finally, purely thermal reactions,
without the help of energetic processes, may well be quite common
\citep{sch93, bos09, the13}.  The formation of NH$_4^+$ and OCN$^-$,
(proposed) carriers of the 6.85 and 4.62 \mum\ features, proceeds in
fact by an acid-base reaction of NH$_3$ with HNCO at temperatures as
low as 15 K \citep{rau04b, bro04}. This is consistent with NH$_4^+$
being the carrier of the 6.85 \mum\ feature, as it is observed in
quiescent sightlines.  HNCO cannot be the only acid involved in this
reaction, however, as the OCN$^-$ abundance is an order of magnitude
below that of NH$_4^+$.

\subsection{Energetic Processing}~\label{sec:modelen}

Historically, it has long been held that the energetic processing of
ices in space plays a key role in the production of new ice species
and specifically in the transformation from simple to more complex
components (e.g., \citealt{gre73}).  Ices in dense clouds experience
CR fluxes that are represented by 1-MeV protons with a flux of 1
proton cm$^{-2}$ $\rm s^{-1}$ (\citealt{moo01} and references
therein). The interstellar UV field of $\sim$ 10$^9$ $G_{\rm 0}$ eV
cm$^{-2}$ s$^{-1}$ (with $G_{\rm 0}=1$ in the local ISM) rapidly
declines with depth into the cloud, regulating the on-set of ice
formation (e.g., \citealt{hol09}). Within clouds, the fluxes of UV
photons are much lower, generated by CR destruction of H$_2$ and
corresponding to $\sim $ 10$^4$ eV cm$^{-2}$ s$^{-1}$. In regions of
star formation these fluxes may be strongly enhanced in both space and
in time.

A vast literature of laboratory studies of energetic processing
exists, which has been particularly successful in solar system studies
(e.g., \citealt{par09}).  In contrast to grain surface chemistry and
thermal processing, the significance of energetic processing of ices
outside the solar system has not yet been unambiguously demonstrated,
however.  Observational tests of energetic processing must address the
similarities and differences between laboratory and observed
interstellar abundances, as well as the spectral profiles of the
produced species of interest and all subsequent chemical
(by-)products.

Many ice absorption features have been studied in the context of
energetic processing, but most attention was paid to those of CO$_2$,
as well as those at 4.62 and 6.85 \mum. CO$_2$ is formed readily by
the energetic processing of most carbon-bearing molecules in the
presence of H$_2$O, including pure H$_2$O ice mantles on carbon grains
\citep{men04}.  \citet{iop13} have explored this to the full extent by
fitting the 15.2 \mum\ CO$_2$ bending mode toward MYSOs, LYSOs, and
dense clouds using only CO$_2$ produced by processing with Galactic
CRs. Comparing these results to the empirical decomposition
(Fig.~\ref{f:15um}), the broad polar component is fitted with CO$_2$
produced from irradiated H$_2$O on carbon grains, the shoulder from
irradiated CH$_3$OH, the apolar components from irradiated CO and
CO:N$_2$ ices, and the pure CO$_2$ component by heating of these
ices. At the observed abundances of CO, H$_2$O, and CH$_3$OH, the
CO$_2$ profiles and abundances are reproduced after $\sim 10^7$ yr for
most targets. Similarly long time scales were found for the production
of CO$_2$ by the CR-generated UV field (see above) in dense clouds
\citep{whi98}. Thus, overall, after the initial rapid formation of
CO$_2$ and H$_2$O by grain-surface reactions (in about $10^5$ yr; see
\S\ref{sec:modelacc}), additional CO$_2$ may be formed by particle
bombardment during the millions of years in YSO environments, or on
shorter time scales for regions exposed to stronger UV and particle
fields. Not all implications of this model have been fully
investigated, however. Most importantly, the effect of a suite of
by-products, in particular carbon-chain oxides (C$_3$O, C$_3$O$_2$;
\citealt{ger96, ger01a, pal08}) on other spectral regions has not yet
been fully assessed. Also, no distinct enhancements of CO$_2$
abundances were observed toward YSOs (Fig.~\ref{f:co2h2o}), although
the situation is complicated by the susceptibility of CO$_2$ itself to
destruction by irradiation.

The 4.62 \mum\ XCN feature was seen for a long time as an indicator of
energetic processing, as laboratory UV photolysis and ion bombardment
experiments on mixed CO and NH$_3$ ices successfully reproduced this
absorption feature with OCN$^-$ (\S\ref{sec:iden}).  Similar
experiments claimed an identification of the 6.85 \mum\ feature with
NH$_4^+$ \citep{sch03} and H$_2$NCONH$_2$ (urea;
\citealt{rau04a}). Both features can also be reproduced through purely
thermal acid-base reactions in ices, however and in many sightlines
this is the more likely process (\S\ref{sec:modeltherm}).

\subsection{Connection with Cometary Ices}~\label{sec:modelcom}

The survival of the protostellar envelope ices upon accretion into the
protoplanetary disk is an open question. Depending on the radius to
the central star, full or partial destruction of ices and subsequent
reformation may have occurred (\citealt{mum11}, and references
therein). The Stardust mission revealed that refractory dust may have
formed close to the early Sun, moving radially outward, accumulating
ices along the way and thus diluting any presolar components in the
outer, comet formation zones \citep{bro14}.  Ice observations provide
constraints to this discussion, but have not been conclusive.  There
are distinct differences between the composition of cometary and LYSO
envelope ices (\S\ref{sec:abunss}). C- and N-bearing species in comets
are under-abundant on average. For C, the abundance distributions do
overlap, which may imply that the early solar system resembled an LYSO
with less C in the ices \citep{obe11}.  Low cometary CO and CO$_2$ may
also reflect sublimation of the most volatile, apolar components upon
accretion from the envelope to the protoplanetary disk \citep{vis09}.
Neither argument holds for N, however: the NH$_3$ abundance
distributions do not overlap and NH$_3$ is only present in the polar
(least volatile) ices.  Direct observations of ices in the disk
mid-plane, where comets are formed, are not possible. Ices in the disk
surface may be linked to the mid-plane by turbulence, however. High
spatial resolution observations have shown that these ices experience
thermal processing (\S\ref{sec:disks}). Further observations are
needed to search for evidence of energetic processing, but chemical
models including vertical disk turbulence have indicated that the ice
composition throughout the disk may be significantly affected by
energetic processes in the disk surface \citep{cie12}.  Radial
transport might lead to such effects as well.

\section{SUMMARY POINTS}~\label{sec:summary}

Infrared observations have revealed the icy Universe
(Fig.~\ref{f:mantle}):

\begin{enumerate}

\item H$_2$O-based ices trace dense clouds ($n\geq 10^3$ \cubcm) at
  depths corresponding to \av$\geq 1.6$ and temperatures below 90 K.
  These are mixed ices, containing at least CO$_2$, NH$_3$, CH$_4$,
  and the carrier of the 6.85 \mum\ feature (NH$_4^+$?).

\item CO freeze-out occurs deeper in dense clouds (\av$\geq 3$),
  forming an apolar ice layer, and it does so ``catastrophically'' at
  \av$\geq 9$ and $n\geq 10^5$ \cubcm, initiating a CO-based chemistry
  on the ices forming a third, CH$_3$OH-rich, layer including also
  H$_2$CO, newly formed CO$_2$, and possibly HNCO.

\item Heating of the ices is commonly observed and the spectroscopic
  signatures provide diagnostics of the thermal history of dense
  environments. Crystallization signatures in the 3.0, 45, and 63
  \mum\ H$_2$O bands, and the position of the 6.85 \mum\ feature trace
  temperatures $\geq 77$ K. The double-peaked 15.2 \mum\ CO$_2$ band
  traces a combination of such high temperatures (``segregation'') and
  temperatures $\geq$20 K as a result of CO sublimation
  (``distillation''). The ratio of the apolar/polar components of the
  4.67 \mum\ CO band itself traces the latter conditions as well.

\item Most observed ice features are consistent with molecule
  formation on cold grain surfaces. The observed ice heating may well
  induce chemical reactions, possibly involving radicals created by
  energetic processes over long time scales. The observational
  evidence for this, possibly features in the 5-7 \mum\ region, is
  weak, however.

\item Variations of the CO$_2$/H$_2$O column density ratios and CO$_2$
  ice composition in local versus GC and extragalactic environments
  likely reflect metallicity differences.

\item Median abundances of C and N-bearing cometary ices are below
  those of the envelopes of LYSOs. The abundance distributions do
  overlap, however, and inheritance of cometary ices from the ISM and
  envelope phases can thus not be excluded.

\item Laboratory spectroscopy has been crucial in the identification
  and in the determination of the chemical and physical history of the
  observed ice features.

\end{enumerate}

\section{FUTURE ISSUES}~\label{sec:future}

\begin{enumerate}

\item Recent high spatial resolution observations of circumstellar
  disks have shown that the ice composition and processing history can
  be traced on scales relevant for planet and comet formation, both
  spectroscopically (e.g., \citealt{sch10, ter12a}) and using
  scattered light imaging (e.g., \citealt{hon09, deb08}). Future high
  angular resolution ground (Thirty Meter Telescope, Extremely Large
  Telescope) and space-based (James Webb Space Telescope; JWST)
  infrared facilities thus promise great advances of ices studies at
  these scales.

\item The same new facilities will enable ice studies in extreme
  environments (strong radiation fields, high or low metallicities),
  such as Galactic and extragalactic MYSOs at larger distances and in
  crowded fields.  Observed feature profile and depth variations may
  further constrain identifications as well as molecule formation
  pathways, including the much sought after evidence for energetic
  processing.

\item The increased sensitivity of new facilities will drastically
  enhance studies of quiescent clouds at very low and high
  extinctions, which are needed to study the ice formation process and
  its interplay with the grain growth process.

\item The continuous 1-29 \mum\ wavelength coverage provided by JWST
  at spectral resolving powers $R$ of up to 3000 will provide strong
  constraints on the identification of new ice species. Supporting
  laboratory experiments are essential for this.

\item New instruments on platforms such as the Stratospheric
  Observatory for Infrared Astronomy (SOFIA) and the Space Infrared
  Telescope for Cosmology and Astrophysics (SPICA) are needed to give
  access to the ice lattice modes ($>25$ \mum), which are particularly
  powerful tracers of the ice composition and thermal history. As they
  appear in emission, the ices may be mapped.

\item The good understanding of the main ice bands allows them to be
  used as tools in the characterization of new sources that will be
  found with new infrared facilities.

\end{enumerate}

\section{RELATED RESOURCES}

Infrared transmission spectra and optical constants of astrophysically
relevant pure ices and mixtures may be found at the following web
sites:

\begin{itemize}

 \item Astrophysical and Astrochemistry Laboratory at NASA Ames
   Research Center:\\ \verb*#http://astrochem.org/db.php#

 \item Experimental Astrophysics Laboratory at Catania Astrophysical
   Observatory:\\
   \verb*#http://www.ct.astro.it/lasp/#

 \item Sackler Laboratory for Astrophysics (Leiden):\\
   \verb#http://www.laboratory-astrophysics.eu/#

 \item Cosmic Ice Laboratory at NASA Goddard Space Flight
   Center:\\ \verb*#http://science.gsfc.nasa.gov/691/cosmicice/#

 \item Grenoble Astrophysics and Planetology Solid Spectroscopy and
   Thermodynamics database (GhoSST):\\
   \verb*#http://ghosst.osug.fr#

 \item Jena - St.Petersburg Database of Optical Constants (mostly
   references to laboratory spectra):\\
   \verb*#http://www.astro.uni-jena.de/Laboratory/Database/jpdoc#

 \item Virtual Atomic and Molecular Data Centre (VAMDC):\\ 
   \verb*#http://portal.vamdc.org#

\end{itemize}

Computer programs for calculating the effect of grain shape and size
on ice absorption band profiles can be found in \citet{boh83} and are
also made available by B. T. Draine at\\

\verb*#http://www.astro.princeton.edu/~draine/scattering.html#

\section{ACKNOWLEDGMENTS}

We thank Deokkeun An, Jean Chiar, Sergio Ioppolo, Karin {\"O}berg,
Joana Oliveira, Laurent Pagani, Elisabetta Palumbo, Klaus Pontoppidan,
Charles Poteet, Gianni Strazzulla, Xander Tielens, and the reviewer
Ewine van Dishoeck for discussions and comments. We also thank
Deokkeun An, Jean Chiar, Emmanuel Dartois, Leslie Looney, Joana
Oliveira, Klaus Pontoppidan, Charles Poteet, Anna Sajina, Jonathan
Seale, and Hiroshi Terada for kindly providing published data in
electronic form.  A.C.A.B. thanks encouragement from Mahsan, Cyrus,
and Jasper to work on this review in a time of moves and job
uncertainty.

\end{document}